\documentclass[lettersize,journal]{IEEEtran}
\usepackage{amsmath,amsfonts}
\usepackage{algorithmic}
\usepackage{algorithm}
\usepackage{array}
\usepackage[caption=false,font=footnotesize,labelfont=rm,textfont=rm]{subfig}
\usepackage{textcomp}
\usepackage{stfloats}
\usepackage[hyphens]{url}
\usepackage{verbatim}
\usepackage{graphicx}
\graphicspath{{figures/}}
\usepackage{cite}
\usepackage{longtable}

\usepackage{hyperref}

\usepackage{pifont}
\usepackage{xcolor}
\usepackage{makecell}
\usepackage[table,xcdraw]{xcolor}
\usepackage{hhline}

\usepackage{multirow}
\usepackage{xcolor,color,framed}
\usepackage[numbers]{natbib}

\usepackage{enumitem}

\begin{document}

\title{Tutorial on Large Language Model-Enhanced Reinforcement Learning for Wireless Networks}

\author{Lingyi Cai, Wenjie Fu, Yuxi Huang, Ruichen Zhang, Yinqiu Liu, Jiawen Kang, Zehui Xiong, \\ Tao Jiang,~\IEEEmembership{Fellow,~IEEE},  Dusit Niyato,~\IEEEmembership{Fellow,~IEEE}, Xianbin Wang,~\IEEEmembership{Fellow,~IEEE},\\ Shiwen Mao,~\IEEEmembership{Fellow,~IEEE}, and Xuemin Shen,~\IEEEmembership{Fellow,~IEEE}

\thanks{Lingyi Cai is with the Research Center of 6G Mobile Communications, School of Cyber Science and Engineering, Huazhong University of Science and Technology, Wuhan, 430074, China, and also with
the College of Computing and Data Science, Nanyang Technological University, Singapore (e-mail: lingyicai@hust.edu.cn).}
\thanks{Wenjie Fu, Yuxi Huang, and Tao Jiang are with the Research Center of 6G Mobile Communications, School of Cyber Science and Engineering, Huazhong University of Science and Technology, Wuhan, 430074, China (e-mail: wjfu99@outlook.com; huangyuxi@hust.edu.cn; tao.jiang@ieee.org).}
\thanks{Ruichen Zhang, Yinqiu Liu, and Dusit Niyato are with the College of Computing and Data Science, Nanyang Technological University, Singapore (e-mails: ruichen.zhang@ntu.edu.sg; yinqiu001@e.ntu.edu.sg; dniyato@ntu.edu.sg).}
\thanks{J. Kang is with the School of Automation, Guangdong University of Technology, Guangzhou 510006, China (e-mail: kavinkang@gdut.edu.cn).}
\thanks{Zehui Xiong is with the School of Electronics, Electrical Engineering and Computer Science (EEECS), Queen’s University Belfast, Belfast BT7 1NN, U.K. (z.xiong@qub.ac.uk).}
\thanks{Xianbin Wang is with the Department of Electrical and Computer Engineering, Western University, London, ON, N6A 5B9, Canada (e-mail: xianbin.wang@uwo.ca).}
\thanks{Shiwen Mao is with the Department of Electrical and Computer Engineering, Auburn University, Auburn, USA (e-mail: smao@ieee.org).}

\thanks{Xuemin Shen is with the Department of Electrical and Computer Engineering, University of Waterloo, Waterloo, ON N2L 3G1, Canada (e-mail: sshen@uwaterloo.ca).}

\thanks{\textit{Lingyi Cai, Wenjie Fu, Yuxi Huang, Ruichen Zhang, Yinqiu Liu contributed equally to this work. (Corresponding author: Tao Jiang.)}}


}

\maketitle

\begin{abstract}

Reinforcement Learning (RL) has shown remarkable success in enabling adaptive and data-driven optimization for various applications in wireless networks. However, classical RL suffers from limitations in generalization, learning feedback, interpretability, and sample efficiency in dynamic wireless environments. Large Language Models (LLMs) have emerged as a transformative Artificial Intelligence (AI) paradigm with exceptional capabilities in knowledge generalization, contextual reasoning, and interactive generation, which have demonstrated strong potential to enhance classical RL. This paper serves as a comprehensive tutorial on LLM-enhanced RL for wireless networks. We propose a taxonomy to categorize the roles of LLMs into four critical functions: state perceiver, reward designer, decision-maker, and generator. Then, we review existing studies exploring how each role of LLMs enhances different stages of the RL pipeline. Moreover, we provide a series of case studies to illustrate how to design and apply LLM-enhanced RL in low-altitude economy networking, vehicular networks, and space–air–ground integrated networks. Finally, we conclude with a discussion on potential future directions for LLM-enhanced RL and offer insights into its future development in wireless networks.

\end{abstract}

\begin{IEEEkeywords}
Reinforcement learning, large language models, LLM-enhanced RL, decision optimization, wireless networks
\end{IEEEkeywords}


\section{Introduction} 


\subsection{Background}


\IEEEPARstart{L}{arge} Language Models (LLMs) have marked a pivotal breakthrough in Artificial Intelligence (AI), significantly reshaping the way machines understand and generate human language \cite{achiam2023gpt,10614634,10679152}. Different from earlier natural language processing models that relied on limited rule-based or shallow learning architectures, LLMs are built upon deep neural networks with hundreds of billions of parameters and trained on vast and diverse corpora of text data \cite{shen2025llm}. Thus, LLMs possess the ability to understand diverse inputs, follow complex instructions, and generate contextually appropriate responses across a wide range of tasks and modalities \cite{NEURIPS2020_1457c0d6,achiam2023gpt,10614634}.

\begin{figure}[t]
\centering
\includegraphics[width=1.01\linewidth]{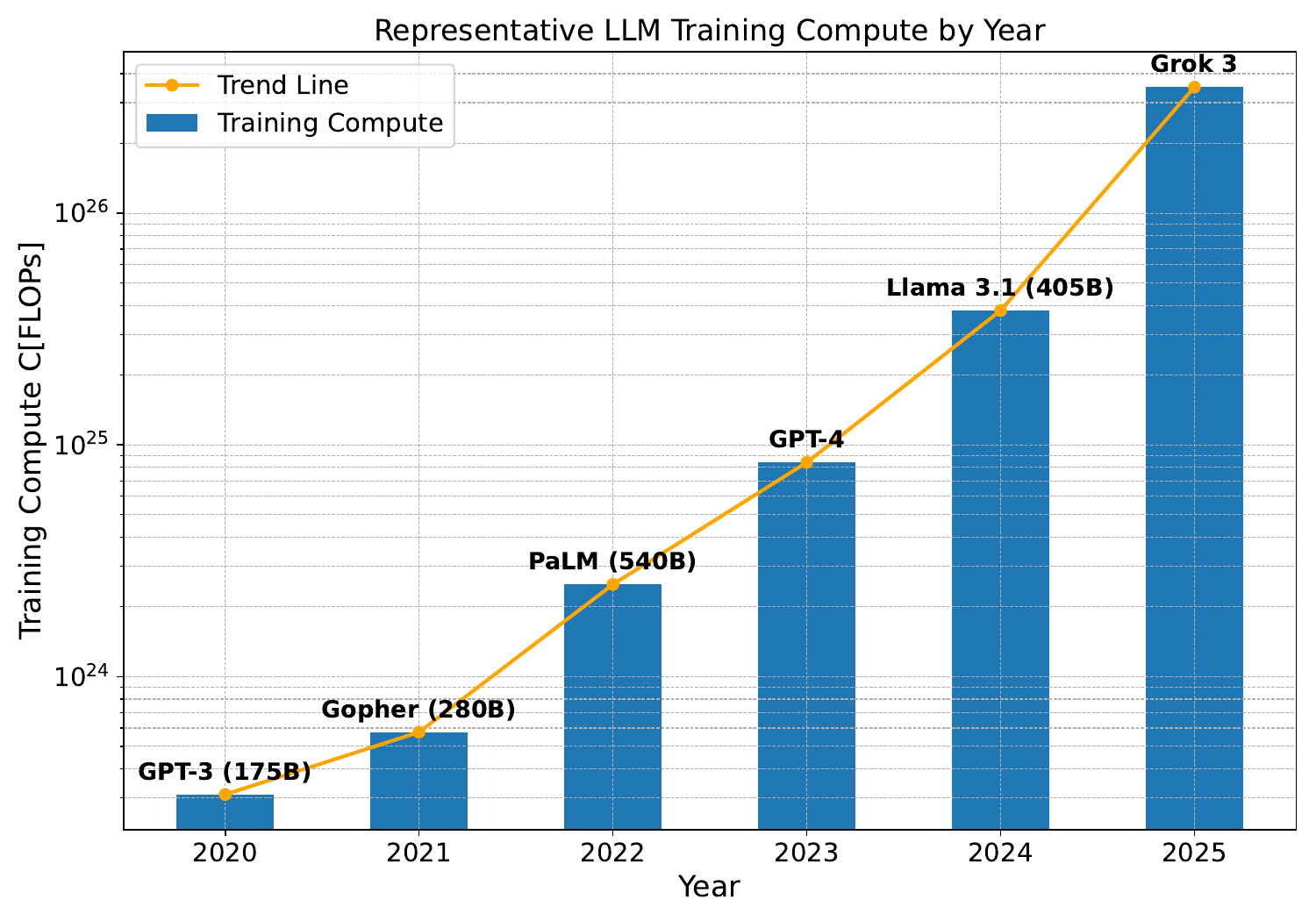}
\caption{Approximate peak training compute requirements (C, measured in floating-point operations (FLOPs)) for one representative LLM each year from 2020 to 2025. The y-axis is log-scaled (base 10), enabling comparison across orders of magnitude. Model names and parameter counts (in billions) are annotated on the bars.}
\label{llmcompute}
\end{figure}

The transformative potential of LLMs has extended to a wide range of applications from conversational agents and code generation tools to knowledge engines and automated reasoning systems. For example, OpenAI’s GPT series\footnote{\label{gpt}\url{https://platform.openai.com/docs/models}}, Google’s Gemini\footnote{\label{palm}\url{https://deepmind.google/models/}}, and Meta’s LLaMA\footnote{\label{llama}\url{https://www.llama.com/models/llama-4/}} models have demonstrated state-of-the-art performance across benchmarks such as question answering, summarization, and reasoning tasks. Moreover, GPT-4 can integrate multi-frame visual input with textual reasoning to enable interpretable and end-to-end driving systems \cite{achiam2023gpt}. Beyond language-centric tasks, LLMs have also begun to play a central role in embodied AI. DeepMind’s Gemini Robotics\footnote{\label{robots}\url{https://deepmind.google/discover/blog/gemini-robotics-brings-ai-into-the-physical-world/?utm_source=chatgpt.com}} equips humanoid robots with LLM-guided control capabilities, enabling them to perform complex real-world tasks such as folding origami and organizing objects. These models differ in terms of scale, adaptation methods, and capabilities, and their evolution has fueled the rapid development of the LLM ecosystem. Fig. \ref{llmcompute} illustrates the trajectory of LLM development through the perspective of training compute \cite{epoch_ml_trends,maslej2025artificial}. The continued rise shows roughly a four- to five-fold increase per year in the training compute. The work in \cite{10.5555/3600270.3602446} demonstrated that performance scales with model size, dataset size, and compute according to power-law relationships. By 2025, models like Grok-3 reflect a leap of over two orders of magnitude in compute compared to GPT-3, signaling the intensifying scale of LLM development.

Overall, the widespread adoption of LLMs is underpinned by several core advantages. They are able to recognize structure across a wide range of inputs and adapt their responses to novel or evolving situations, where strength lies not only in recognizing patterns but in modeling intent, evolving with context, and supporting coherent decision-making \cite{10778660}. Their ability to model context, incorporate prior knowledge, and refine behavior through iterative feedback makes them particularly well-suited for dynamic problem-solving \cite{NEURIPS2023_91edff07}. Furthermore, their capacity to simulate, organize, and synthesize complex relationships allows them to accelerate learning and coordination across diverse tasks \cite{cai2025large}. These capabilities collectively position LLMs as a foundation and engine in increasingly dynamic and intelligent systems.

\subsection{Motivations}

Reinforcement learning (RL) has demonstrated considerable promise in the operation of wireless networks by enabling data-driven and adaptive solutions to complex decision-making problems \cite{9372298,10283826,9195488}. It has been successfully applied to a variety of core network optimization tasks, including dynamic spectrum access \cite{10415879}, power control \cite{9539879}, resource scheduling \cite{10697115,10623528}, and UAV trajectory planning \cite{10422956}. These applications often involve high-dimensional state spaces, time-varying dynamics, and conflicting performance objectives, in which RL demonstrates its strength by learning optimal policies through interaction with the environment \cite{9372298,9575181}. However, as wireless systems evolve toward increasing complexity, scale, and heterogeneity, classical RL methods face four growing limitations: (i) generalization and multimodal understanding, (ii) feedback bottlenecks in learning signals, (iii) decision instability and lack of interpretability, and (iv) sample inefficiency and multi-task inadaptability, which are discussed in detail later in Section \ref{limitationsofrl}.



In the context of the recent rise of LLMs, new possibilities have emerged to address the abovementioned challenges of RL in wireless systems. Thanks to the remarkable abilities (as detailed in Section \ref{LLMabilities}) on knowledge generalization, contextual reasoning, and interactive generation \cite{cai2025large,10778660,11059888}, LLMs have evolved from text generators to universal cognitive engines that can model, simulate, explain, and guide complex decision processes \cite{11175216,10670196,10380515}. The success in domains such as robotics\textsuperscript{\ref {robots}} and human–AI interactionT\footnote{\label{AutoGPT}\url{https://github.com/Significant-Gravitas/AutoGPT}}  suggests that LLMs could offer opportunities to complement the weaknesses of classical RL. Therefore, the integration of LLMs into RL presents a compelling paradigm shift. By leveraging the strengths of LLMs, this integration can offer a high-level semantic interface that enables the agent to incorporate task semantics, contextual understanding, and domain-specific priors into the learning process, thereby shaping learning beyond raw experience \cite{xie2024textreward,ma2024eureka,10531073}. Thus, in addition to relying on learning through environment interaction, RL agents can be enhanced in terms of understanding task intent, simulate plausible outcomes, explain actions, and accelerate learning \cite{10766898,cai2025large,10945973}. The enhancement is particularly promising for wireless systems, where expert knowledge is hard to encode, real-world feedback is expensive, and decision-making is often difficult due to trade-offs in network optimization such as balancing latency, energy consumption, and throughput \cite{8714026,9372298,10283826}.

To this end, this tutorial adopts a structured viewpoint and proposes a taxonomy of LLM-enhanced RL frameworks tailored for wireless networks. We classify the roles of LLMs into four critical functions, state perceiver, reward designer, decision-maker, and generator, to analyze how each role enhances different stages of RL. This perspective provides a unified framework for understanding, designing, and deploying the LLM-enhanced RL framework, which integrates prior knowledge, contextual reasoning, and generative capabilities to support intelligent learning and decision-making to meet the complexity and diversity of future wireless networks.

\begin{table}[t]
\centering
\caption{SUMMARY OF RELATED SURVEYS}
\label{tab:relevant_surveys}
\begin{tabular}{|c|p{0.36\textwidth}|}
\hline
\multicolumn{1}{|c|}{\textbf{Reference}} & \multicolumn{1}{c|}{\textbf{Focus}} \\
\hline
\cite{8714026} & A survey that systematically maps DRL algorithms to core wireless networking tasks, establishing an MDP-based framework for intelligent and adaptive network control \\
\hline
\cite{9201129} & A survey that reviews RL applications from a layer-wise perspective of wireless networks, highlighting cross-layer optimization, algorithm–task alignment, and challenges in building unified and interpretable learning frameworks \\
\hline
\cite{9403369} & A review that provides a comprehensive analysis of DRL in IoT systems, introducing a unified framework for communication, computing, caching, and control \\
\hline
\cite{9372298} & A tutorial that unifies single-agent, multi-agent, and model-based DRL for AI-enabled wireless networks, emphasizing cooperative and distributed learning paradigms for scalable and autonomous network optimization \\
\hline
\cite{9738819} & A survey that provides a comprehensive overview of multi-agent RL for future Internet architectures, highlighting Markov game formulations, cooperative and competitive learning algorithms, and applications in distributed and autonomous network management \\
\hline
\cite{10283826} & A survey that reviews reinforcement learning-based approaches for autonomous multi-UAV wireless networks, which emphasize perception, decision, and action frameworks to enable key intelligent tasks such as efficient communication and adaptive trajectory optimization  \\
\hline
\end{tabular}
\end{table}

\subsection{Related Work}

In recent years, many surveys and tutorials have emerged to capture the evolution of RL techniques in wireless networks from single-agent DRL to multi-agent DRL applied to communication systems and intelligent networks. By enabling autonomous agents to learn optimal control strategies through interaction with the environment, RL provides a powerful framework for addressing the complexity and heterogeneity of modern wireless networks.

\begin{figure*}[!t]
\centering
\includegraphics[width=5.5in]{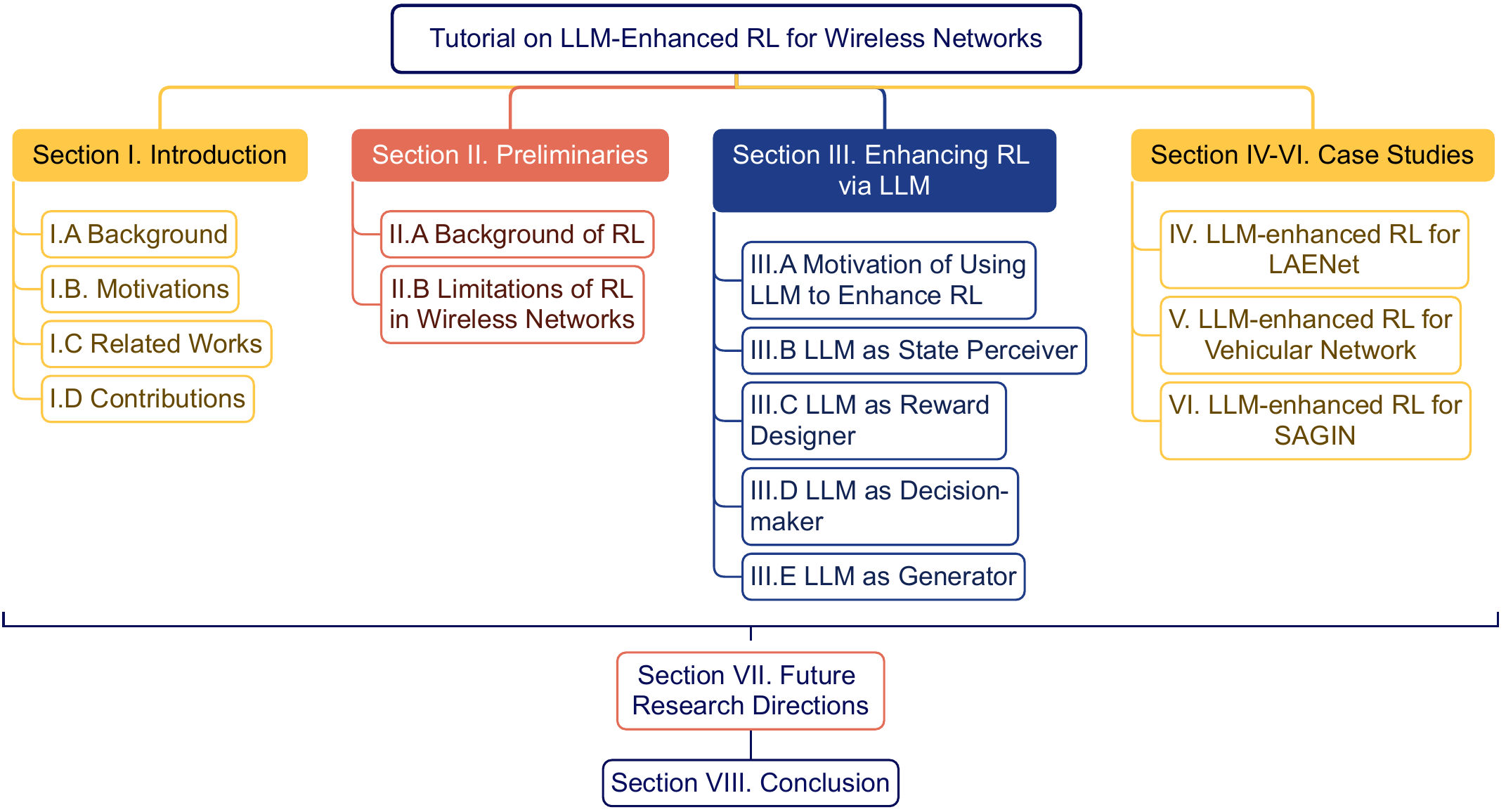}
\caption{Overall structure of the tutorial on LLM-enhanced reinforcement learning (RL) for wireless networks. The tutorial is organized into eight sections, beginning with the introduction and preliminaries, followed by a taxonomy of LLM roles in enhancing RL, including the functions of state perceiver, reward designer, decision-maker, and generator. Subsequent sections present case studies in representative network scenarios, while the final sections discuss future research directions and conclude the tutorial.}
\label{organizefig}
\end{figure*}

The work in \cite{8714026} provides a comprehensive survey of DRL applications in communications and networking. It outlines the MDP formulation underlying DRL and emphasizes its advantages in handling large-scale state–action spaces through adaptive decision-making and distributed learning. The paper further summarizes DRL’s progress across key network functions such as dynamic spectrum access, rate control, caching, offloading, routing, and security, offering an integrated view of DRL’s role in intelligent communication systems. The subsequent survey \cite{9201129} adopted a layering view, which categorized RL applications across the physical, medium access control, network, and application layers. It shows how DRL models such as Q-learning and Actor–Critic architectures enable cross-layer control while emphasizing the need for unified decision-making frameworks. The authors in \cite{9403369} provide a comprehensive taxonomy of major DRL algorithms and review their applications in Internet of Things (IoT) domains such as smart grid, intelligent transportation, and mobile crowdsensing. Moreover, the study systematically reveals how DRL can address decision-making and performance optimization challenges in communication, computing, caching, and control without relying on prior system models. Subsequently, the study in \cite{9372298} unified single-agent and multi-agent RL perspectives in the tutorial for AI-enabled wireless networks. They review both model-free and model-based DRL frameworks and further emphasize multi-agent reinforcement learning (MARL) as a key enabler for decentralized and cooperative control. This paper categorizes MARL approaches into learned cooperation, emergent communication, and networked agents, and reviews their applications in mobile edge computing, UAV networks, and cell-free massive MIMO. Furthermore, the work in \cite{9738819} provides an in-depth investigation of MARL and its applications in the future Internet. It systematically classifies MARL algorithms into value decomposition, policy gradient and actor–critic categories, and examines their mechanisms for cooperation, competition, and communication among agents. The paper highlights MARL’s capability to address dynamic, decentralized, and large-scale decision-making problems, offering a comprehensive perspective on how MARL can enable autonomous and adaptive control in next-generation network systems. The authors in \cite{10283826} explore RL from the more specific perspective of autonomous multi-UAV wireless networks. The paper provides a comprehensive overview of how RL techniques enable UAVs to make intelligent and adaptive decisions in dynamic wireless environments
. It categorizes existing approaches into model-free, model-based, and multi-agent RL frameworks, and highlights RL’s advantages in enhancing the autonomy, scalability, and energy efficiency of UAV networks.

The existing related works demonstrate that RL has achieved remarkable success in enabling intelligent and adaptive control for complex wireless networks. These efforts have also advanced the field from single-agent decision-making to cooperative multi-agent optimization, greatly improving resource utilization and network autonomy in wireless systems. However, in essence, the above research still focuses on the paradigm of classical RL, which relies on handcrafted state representations and handcrafted rewards with trial-and-error learning within task-specific environments \cite{9247965,10283826}. As the next generation of wireless networks evolves toward more dynamic, heterogeneous, and human-centric architectures, the classical RL approaches may face fundamental limitations in scalability, generalization, and adaptation \cite{9372298,8807386,10288078}.

Encouragingly, the recent emergence of LLMs offers new opportunities with capabilities of semantic perception, contextual reasoning, and generative world modeling \cite{10766898,10.1145/3641289}. These abilities align naturally with the unmet needs of classical RL in wireless networks, such as understanding complex network semantics \cite{10778660}, designing adaptive states and rewards \cite{wang2025chain}, and interpreting RL agents' behaviors and policies \cite{NEURIPS2023_ae9500c4}. By bridging the powerful capabilities of LLMs with RL, our work advances beyond traditional RL surveys on wireless networks and offers a forward-looking tutorial on how LLMs can enhance classical RL for wireless networks.

\subsection{Contributions}

The main contributions of our tutorial are summarized as follows:

\begin{itemize}

    \item We propose a unified framework that functions with LLMs as state perceiver, reward designer, decision-maker, and generator to enhance different stages of the RL pipeline and review existing efforts on LLM-enhanced RL in wireless networks.

    \item We provide several case studies to demonstrate the practical applications and effectiveness of LLM-enhanced RL in future wireless network scenarios, including low-altitude economy networking, vehicular networks, and space–air–ground integrated networks.

    \item We discuss potential future directions and open challenges for LLM-enhanced RL, which aim to guide the evolution of LLM-enhanced RL toward more adaptive, scalable, secure, interpretable, and intelligent wireless systems.

\end{itemize}

As shown in Fig. \ref{organizefig}, the rest of this tutorial is organized as follows. In Section II, we present the fundamental background of RL and analyze its limitations in dynamic wireless environments. Section III introduces the motivation and key principles of integrating LLMs into RL. Then, a taxonomy categorizes the roles of LLMs into four critical roles and reviews existing studies on how to enhance RL via LLM in each role in applications of wireless networks. Sections IV to VI provide several case studies to illustrate how LLM-enhanced RL can be implemented in low-altitude economy networks, vehicular networks, and space–air–ground integrated networks. Finally, Section VII discusses open challenges and future research directions, and Section VIII concludes this tutorial.


\section{Preliminaries} 

\subsection{Background of RL}

\subsubsection{Fundamentals of RL}

\begin{figure}[t]
\centering
\includegraphics[width=0.9\linewidth]{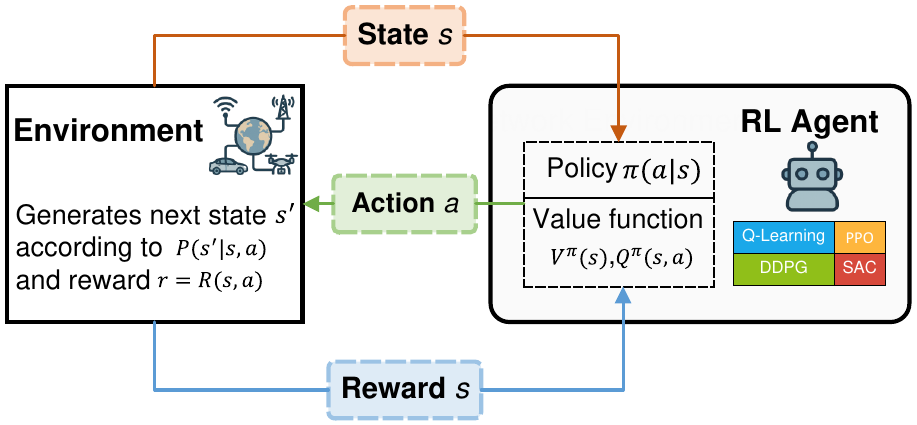}
\caption{The classical RL framework in wireless systems, where the RL agent interacts with the environment by observing the state $s$, selecting an action $a$ according to the policy $\pi(a|s)$, and receiving a reward $r = R(s,a)$. The environment then transitions to the next state $s'$ according to $P(s'|s,a)$, enabling the agent to optimize its value function $V^{\pi}(s)$ or $Q^{\pi}(s,a)$ through iterative learning.}
\label{rlknow}
\end{figure}

As shown in Fig. \ref{rlknow}, RL is a type of machine learning where an agent learns to make decisions by interacting with an environment to maximize long-term rewards. RL can be formalized using the framework of MDPs, which consist of the following components \cite{10032267}:

\begin{enumerate}[label=(\alph*)]
	\item $S$: the set of all states.
	\item $A$: the set of all actions, which can be discrete or continuous.
	\item $P$: the state transition probability function, where $P(s'|s,a)$ represents the probability of transitioning to state $s'$ from state $s$ after taking action $a$.
	\item $R$: the reward function, where $R(s,a)$ gives the immediate reward received after taking action $a$ in state $s$.
	\item $\gamma \in [0, 1]$: the discount factor, which determines the importance of future rewards.
\end{enumerate}

The agent's goal is to learn a policy $\pi(a|s)$ that maps states to action \cite{10304612}. Such a policy can be deterministic (i.e., $\pi: S \rightarrow A$) or stochastic (i.e., $\pi: S \rightarrow p(A|S)$).
The agent interacts with the environment in discrete time steps. At each time step $t$, the agent observes the current state $s_t$, selects an action $a_t$ according to its policy $\pi(a_t|s_t)$, receives a reward $r_t = R(s_t, a_t)$, and transitions to the next state $s_{t+1}$ based on the transition probability $P(s_{t+1}|s_t, a_t)$. The objective of the agent is to maximize the expected cumulative discounted reward, also known as the return:

\[G_t = \sum_{k=0}^{\infty} \gamma^k r_{t+k}.\]

To measure the long-term value of states and actions, RL uses value functions \cite{10506539}. The state-value function $V^{\pi}(s)$ represents the expected return when starting from state $s$ and following policy $\pi$ thereafter:

\[V^{\pi}(s) = \mathbb{E}_{\pi} [G_t | s_t = s].\]

The action-value function $Q^{\pi}(s,a)$ represents the expected return when starting from state $s$, taking action $a$, and following policy $\pi$ thereafter:

\[Q^{\pi}(s,a) = \mathbb{E}_{\pi} [G_t | s_t = s, a_t = a].\]

\subsubsection{Classic RL Algorithms}

RL algorithms can be broadly classified into three main families based on what they learn and optimize:

Value-Based Methods focus on learning an optimal value function, from which a policy is implicitly derived. A canonical example is Q-Learning, a model-free, off-policy algorithm that learns the optimal action-value function, $Q^*(s,a)$, by iteratively updating its estimates based on the Bellman optimality equation~\cite{watkins1992q}. The policy is then typically formed by selecting the action with the highest Q-value in a given state. The advent of deep learning led to the Deep Q-Network (DQN), which employs a neural network to approximate the Q-function, making it applicable to high-dimensional state spaces~\cite{mnih2013playing}. DQN incorporates techniques like experience replay and target networks to stabilize training.

Policy-Based Methods directly learn a parameterized policy, $\pi_\theta(a|s)$, without needing an intermediate value function. These methods optimize the policy parameters $\theta$ by performing gradient ascent on the expected return. They are particularly effective in continuous or high-dimensional action spaces where value-based methods struggle. However, policy gradient estimates can suffer from high variance, which can make training unstable.

Actor-Critic Methods combine the strengths of the previous two approaches by maintaining two separate models: an actor and a critic. The actor, which represents the policy, selects actions, while the critic, which represents the value function, evaluates these actions. The critic's feedback provides a low-variance learning signal to guide the updates of the actor's policy, leading to more stable and efficient learning. This hybrid architecture forms the foundation for many state-of-the-art Deep Reinforcement Learning (DRL) algorithms, including:
\begin{enumerate}[label=(\alph*)]
	\item Deep Deterministic Policy Gradient (DDPG)~\cite{lillicrap2015continuous}: An off-policy actor-critic algorithm designed for continuous action spaces.
	\item Proximal Policy Optimization (PPO)~\cite{schulman2017proximal}: An on-policy algorithm known for its stability and ease of implementation. It prevents large, destructive policy updates by using a clipped surrogate objective function, making it a robust and widely used choice.
	\item Soft Actor-Critic (SAC)~\cite{haarnoja2018soft}: An off-policy actor-critic algorithm that incorporates an entropy maximization term into its objective. This encourages exploration and often leads to more robust and sample-efficient learning in continuous action spaces.
\end{enumerate}

\subsection{Limitations of RL in Wireless Networks}
\label{limitationsofrl}

RL has shown great potential in enabling intelligent decision-making and optimization across various aspects of wireless networks operations. However, the practical application of RL in dynamic wireless environments still faces several critical limitations as follows.

\begin{itemize}
\item {\bf Generalization and Multimodal Understanding}: In wireless network scenarios, RL faces challenges related to limited generalization capability and difficulties in multimodal understanding. Traditional RL methods typically rely on numerical state representations derived from specific environments \cite{9372298}, making it difficult to transfer state observations and action policies to dynamic and heterogeneous network scenarios, such as Open Radio Access Network (O-RAN) slicing \cite{lotfi2025prompt}, UAV networks \cite{dharmalingam2025aero}, and edge computing \cite{s25010175}. Moreover, classical RL agents are usually designed to process single-modality inputs from environments, such as state vectors \cite{10299604}. Emerging network systems increasingly involve multimodal inputs (e.g., natural language instructions, image data, and log texts) \cite{10660494}, which further increases the cognitive burden on RL agents and limits their scalability in practical applications.

\item {\bf Feedback Bottlenecks in Learning Signals}: It is challenging to design effective reward functions (i.e., the most important learning signal feedback) for RL to optimize multiple objectives in wireless networks, such as throughput, energy consumption, latency, and quality of service \cite{9247965}. Improper reward design without appropriate trade-offs among these objectives may introduce bias to the agent toward over-optimizing a single metric (e.g., maximizing throughput at the cost of excessive delay), leading to slow convergence and suboptimal policies. Moreover, the RL agents obtain reward feedback after long operation periods due to the highly dynamic nature of wireless environments and the wide state space, leading to sparse and delayed reward signals that further exacerbate the instability of the policy learning process \cite{10480915,9427224}.

\item {\bf Decision Instability and Lack of Interpretability}: On the one hand, model-based RL methods rely on accurate environment modeling. However, wireless environments are highly dynamic and uncertain due to factors such as unpredictable user mobility and time-varying channel conditions \cite{8807386,10283826}. In such cases, model errors can easily accumulate, leading to failures in policy learning such as resource allocation errors or service interruptions. On the other hand, the decision-making process of traditional DRL often operates as a closed-box mechanism, lacking transparency and interpretability to provide clear causal reasoning or decision rationales \cite{10.1145/3527448,10.1145/3616864,10.1145/3623377}. These limitations may prevent systems from passing regulatory audits or gaining user trust, thereby posing significant risks in high-stakes scenarios such as 6G network slicing, UAV scheduling, and Vehicle-to-Infrastructure (V2I) communications.

\item {\bf Sample Inefficiency and Multi-task Inadaptability}: On the one hand, classical RL relies on interactions with the environment to collect extensive training data \cite{10288078}. However, in real-world wireless scenarios, each interaction may incur significant costs (e.g., energy consumption and time delays) or pose potential risks such as communication failures, leading to inefficiency and high deployment costs in training processes. On the other hand, wireless network often handle multiple tasks simultaneously such as channel estimation and computing offloading \cite{9079564}. Traditional RL is usually task-specific and lacks abilities for knowledge transfer, making it difficult to share experience or formulate policy models across tasks \cite{10614179}. As a result, even small changes in tasks (e.g., resource allocation or QoS assurance) often necessitate retraining from scratch.

\end{itemize}

The challenges discussed above highlight the limitations of classical RL when applied to wireless networks. Fortunately, recent advances in integrating LLMs with RL offer promising directions for addressing these challenges. In the following section, we will explore how the LLMs are used to enhance the RL for wireless networks.






\section{Enhancing RL via LLM} 

\subsection{Motivation of Using LLM to Enhance RL} 
\label{LLMabilities}

\begin{figure}[t]
	\centering
	\includegraphics[width=\linewidth]{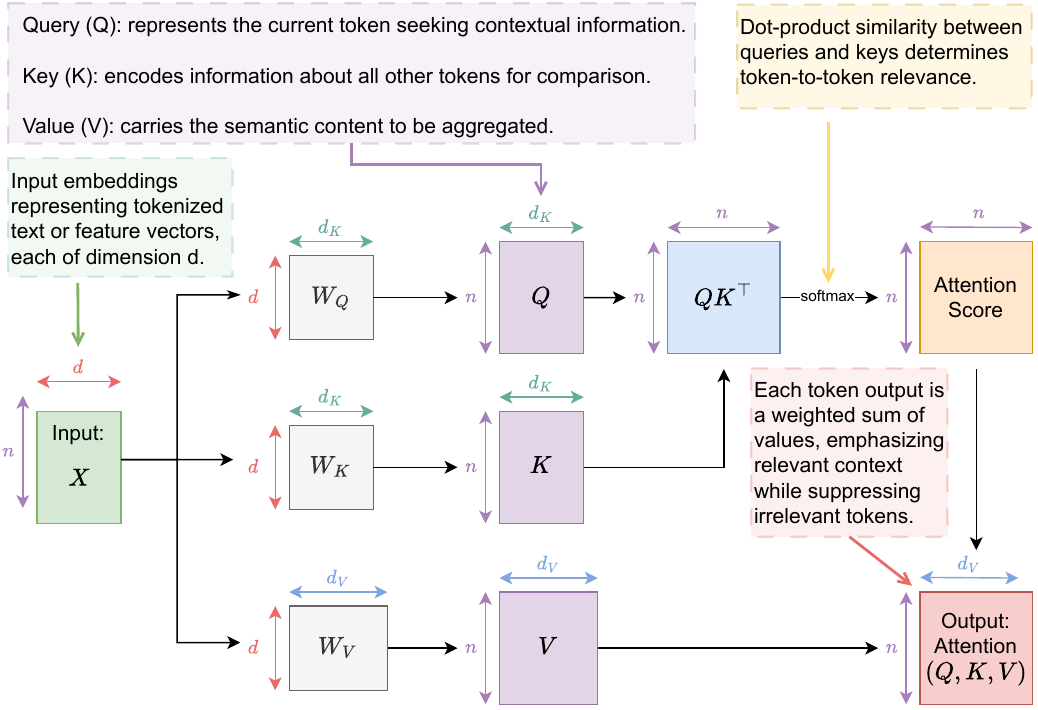}
	\caption{Self-attention mechanism in Transformer architecture. The input sequence is first projected into query, key, and value representations through learned linear mappings. Dot-product similarity between queries and keys determines token-to-token relevance, which is normalized via softmax to produce attention weights. These weights form a weighted aggregation of value vectors, enabling each token to integrate contextual information from all others and produce a meaning-aware output representation.}
	\label{fig:self-attention}
\end{figure}

\subsubsection{Background of LLM}
Generally, language models are trained to predict the next word in a given text sequence, which enables them to learn complex patterns and relationships in language.
To further enhance their capabilities in understanding, generating, and reasoning over natural language, LLMs are developed by training advanced neural network architectures with billions of parameters on massive corpus datasets~\cite{longpr2024Pretrainer,10879580}.
LLMs are typically based on the Transformer architecture, which utilizes self-attention mechanisms to capture long-range dependencies and contextual information in text data~\cite{vaswan2017Attention}.

Mathematically, as shown in Figure~\ref{fig:self-attention},
given an input sequence represented by a matrix of word embeddings $X \in \mathbb{R}^{n \times d}$, where $n$ is the sequence length and $d$ is the dimensionality of the embeddings, the self-attention mechanism first projects $X$ into three matrices: the query $Q$, key $K$, and value $V$:
\[
Q = XW_Q, \quad K = XW_K, \quad V = XW_V,
\]
where $W_Q, W_K \in \mathbb{R}^{d \times d_K}, W_V \in  \mathbb{R}^{d \times d_V}$ are learnable weight matrices, $d_K$ is the dimensionality of the queries and keys, and $d_V$ is the dimensionality of the values.

The attention scores are then computed as the scaled dot-product between the queries and keys:
\[
\text{Attention} (Q, K, V) = \text{softmax} \left(\frac{QK^\top}{\sqrt{d_K}} \right)V,
\]
where the softmax function $\text{softmax}(z_i) = \frac{e^{z_i}}{\sum_{j} e^{z_j}}$ normalizes the scores to obtain a probability distribution over the values.
This formulation ensures that the model computes the relevance of each word in the sequence with respect to every other word, scaled by the factor $\sqrt{d_K}$ to prevent overly large dot product values that could lead to numerical instability.
Stacked with multiple layers of self-attention and other components like feed-forward networks, normalization, and residual connections, the Transformer architecture allows LLMs to effectively capture complex relationships in language data.

\begin{figure*}[!t]
\centering
\includegraphics[width=6.8in]{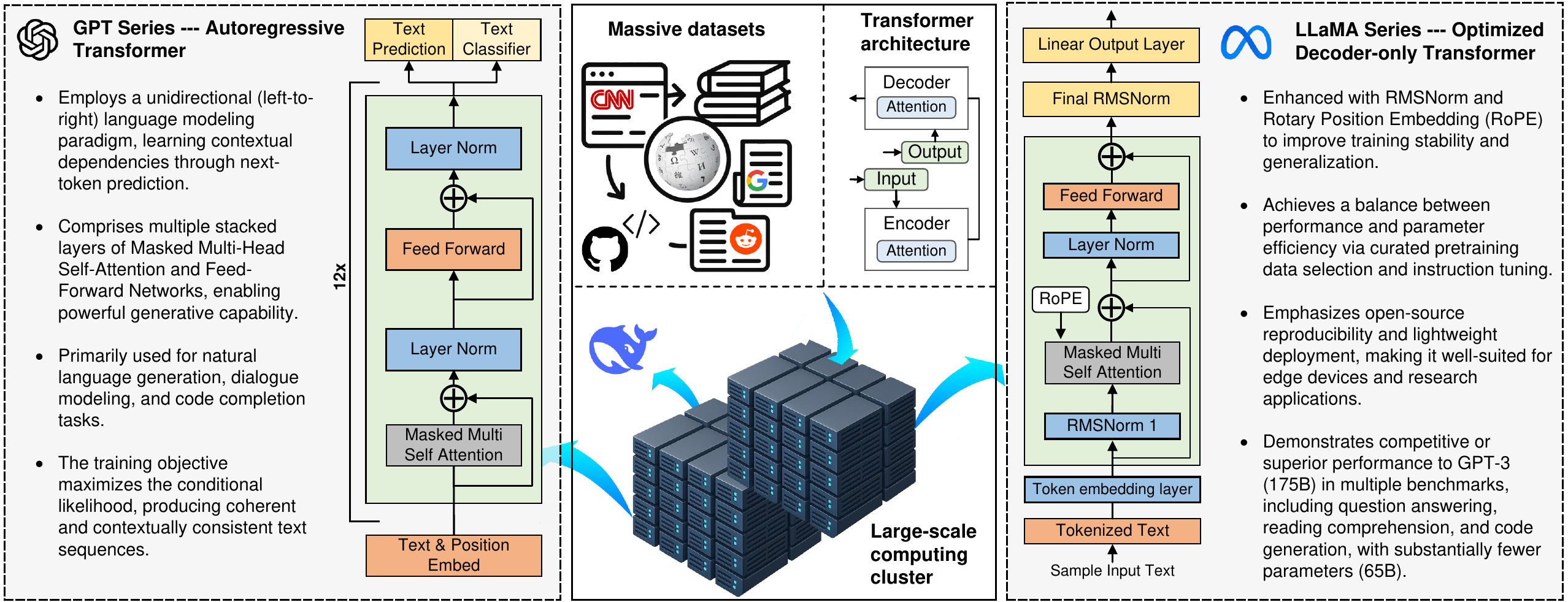}
\caption{Overview of representative LLM architectures and their training paradigms. The GPT series (left) adopts an autoregressive transformer with a unidirectional attention mechanism for next-token prediction, excelling in text generation and dialogue modeling. The central panel illustrates the large-scale pretraining pipeline, where massive multimodal datasets are processed through the transformer architecture. The LLaMA series (right) introduces an optimized decoder-only transformer enhanced with RMSNorm and Rotary Position Embedding (RoPE), achieving improved training stability, parameter efficiency, and adaptability for lightweight or edge-oriented deployment.}
\label{LAENetjiagoutu}
\end{figure*}

GPT (Generative Pre-trained Transformer) models, which advance autoregressive pretraining, enable fluent text generation thorugh next-token prediction~\cite{radfor2018Improving,radforLanguageM,brown2020LanguageM}.
BERT (Bidirectional Encoder Representations from Transformers) models, which introduce bidirectional atthention layers, enable deep contextual understanding by predicting masked tokens in a sequence~\cite{devlin2019BERTPretr}.
Later, models like RoBERTa~\cite{liu2019RoBERTaRo} and T5~\cite{raffel2020Exploring} refined training efficiency and task transferability, and LLaMA~\cite{touvro2023LLaMAOpen,touvro2023Llama2Op,gratta2024Llama3He} demonstrated that carefully curated data and architectural optimizations can achieve strong performance with fewer parameters,
where the first version of LLaMA (65B parameters) achieves most state-of-the-art results than contemporary GPT-3 (175B parameters) in multiple tasks including natural question answering, reading comprehension, and code generation.
These advancements have led to the development of LLMs that can understand, generate, and reason over natural language with remarkable fluency and accuracy not only in research but also further in practical applications and cross domain tasks including reinforcement learning~\cite{cao2025SurveyLar}.

\subsubsection{Key capabilities of LLM} We outline four key capabilities of LLMs for supporting RL  in complex wireless network environments.evolution of LLMs has been driven by key innovations in model design, training objectives, and scaling strategies.

\begin{itemize}

\item {\bf Generalization and Multimodal Comprehension}: LLMs are pretrained on vast and diverse datasets, enabling them to understand and process a wide range of input modalities, such as textual commands, spatial layouts, visual inputs, and structured data \cite{10.1145/3641289,11028927}. This capability of LLMs can assist RL agents to better interpret complex wireless environments and generalize across different tasks and domains. For example, LLMs can be used to parse high-level service requirements expressed in natural language \cite{pesl2025adopting}, allowing the RL agent to better align decisions with user intent and wireless system constraints. Additionally, LLMs can assist agents in better interpreting various elements of wireless networks \cite{10648594}, such as traffic logs or sensor signals, thereby enhancing the agent's ability to make decisions in subsequent stages.

\item {\bf Context-Aware and Reward Shaping}: LLMs possess strong capabilities in semantic understanding, contextual modeling, and domain knowledge integration \cite{10778660}. Thus, they can serve as intelligent reward designers in place of manually crafted or static reward functions traditionally used in RL based on expert knowledge. On the one hand, LLMs can generate reward functions based on optimization objectives \cite{10.1145/3626772.3657767}, system constraints, and environmental states, thereby effectively balancing multiple goals in wireless optimization such as throughput, energy consumption, latency, and QoS. On the other hand, LLMs can evolve reward functions by quickly adapting semantic understanding and abstraction \cite{pmlr-v202-du23f}. As a result, RL agents can leverage adaptive reward shaping in complex and dynamic wireless network environments to guide policy learning, enabling more flexible responses to changing objectives and tasks in wireless networks.

\item {\bf Transparent Reasoning and Stable Decision-Making}: By employing prompt strategies such as Chain-of-Thought (CoT), LLMs can guide RL agents to generate step-by-step reasoning chains to enable more interpretable, logically coherent, and causally grounded decisions \cite{wang2025chain}. This structured reasoning process improves the explainability of RL policies \cite{NEURIPS2022_9d560961}. For example, translating complex decisions such as resource allocation into natural language explanations and supporting the generation of interpretable fault-recovery strategies that comply with domain constraints. Moreover, LLMs exhibit strong capabilities in cross-domain knowledge transfer and memory of past experiences, which helps reduce the possibility of random or suboptimal behaviors \cite{NEURIPS2024_60960ad7}. Such approach facilitates the sharing of key information across different tasks and domains \cite{zhong2024memorybank}, allowing RL agents to maintain policy stability even when faced with new environments or changing conditions in wireless networks.

\item {\bf Sample Generation and Multi-task Management}: Empowered by extensive pre-trained knowledge and strong generalization capabilities, LLMs can generate high-quality synthetic samples through simulation, in-context learning, or prompt engineering \cite{pmlr-v202-nottingham23a,NEURIPS2023_ae9500c4}. These generated samples (e.g., traffic patterns, network topologies, and signal types) help reduce dependence on costly real-world wireless environment interactions, thereby improving RL training efficiency and accelerating policy learning. Additionally, LLMs possess broad comprehension and transfer capabilities to share knowledge across tasks based on their understanding of domain-specific concepts to support multi-task management and coordination \cite{10890371,NEURIPS2023_6b8dfb8c}. For example, the LLM can transfer knowledge of resource allocation gained from a channel estimation task to the optimization of computation offloading strategies by identifying commonalities between them (e.g., bandwidth usage and energy constraints), thus facilitating multi-task optimization of communication and computation resources for RL.

\end{itemize}

To systematically investigate enhancing RL via LLMs in wireless networks, we propose a taxonomy framework that categorizes existing works according to four key roles that LLMs can assume within RL frameworks: state perceiver, reward designer, decision maker, and generator, as shown in Fig. \ref{LAENetjiagoutu}. The following parts examine each role in detail to highlight representative methods, design principles, and application strategies \footnote{\label{fnprompts}The summary of datasets and code repositories for LLM-enhanced RL for wireless networks and other domains is available at: \url{https://github.com/ResearcherSG/Tutorial-on-Large-Language-Model-Enhanced-Reinforcement-Learning-for-Wireless-Networks}} .

\begin{figure*}[!t]
\centering
\includegraphics[width=5.8in]{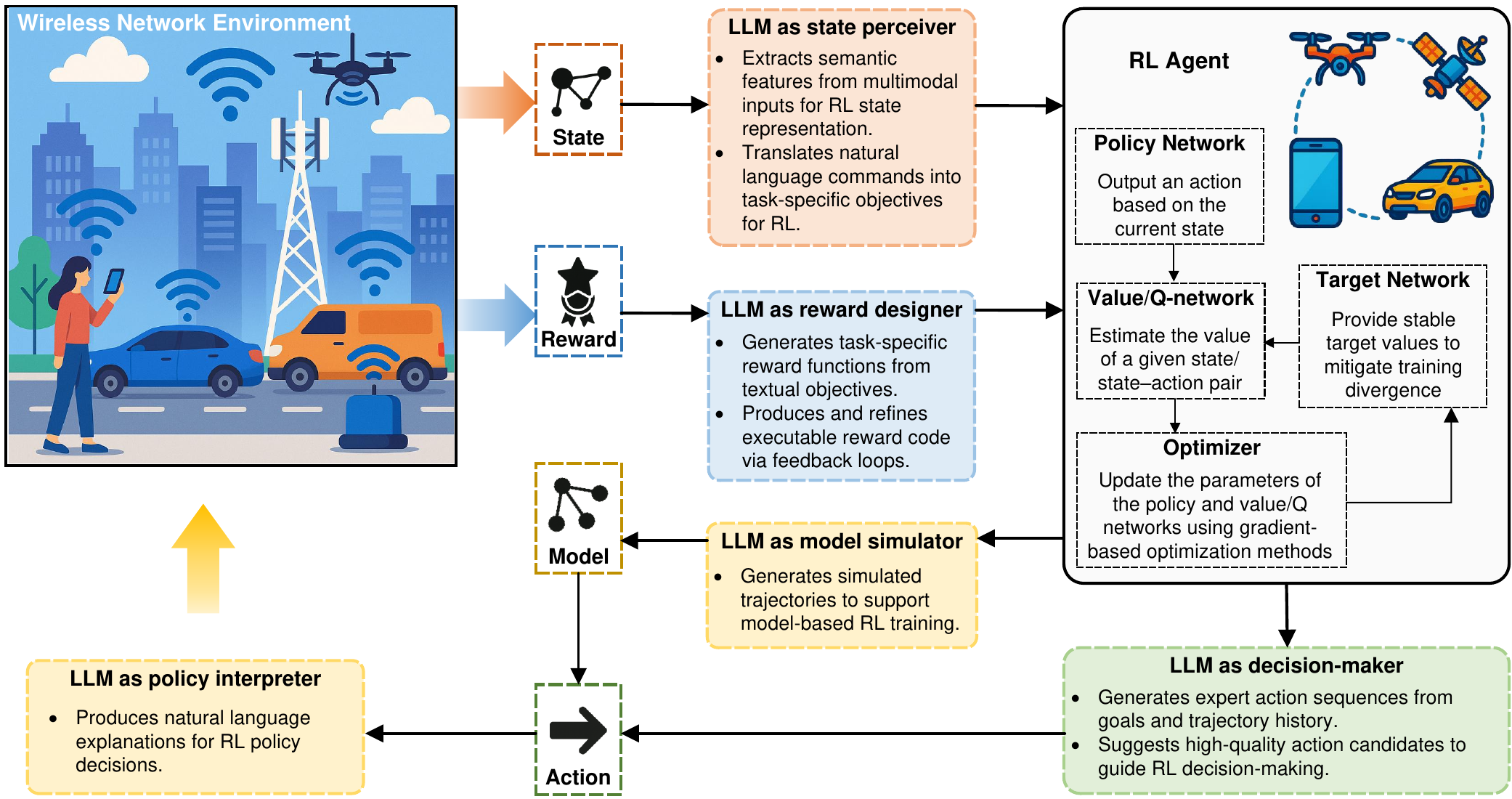}
\caption{The overview and architecture of LLMs in enhancing RL for wireless network environments. LLMs assist as state perceivers, reward designers, generators (including model simulators and policy interpreters), and decision-makers. Key capabilities and functions of each role are illustrated across various network scenarios including MEC, vehicular networks, SAGIN, and UAV networks.}
\label{LAENetjiagoutu}
\end{figure*}

\subsection{LLM as State Perceiver} 

\begin{table*}[] 
\centering 
\caption{Summary of LLMs as state perceivers for Enhancing RL \\ CIRCLES (\textcolor{blue!20}{\ding{108}}) DESCRIBE THE METHODS; CORRECT MARKERS (\textcolor{green}{\ding{51}}) AND CROSS MARKERS (\textcolor{red}{\ding{55}}) REPRESENT PROS AND CONS RESPECTIVELY.} 
\label{tab:my-table} 
\centering 
\begin{tabular}{|c|c|c|l|} 
\hhline{|-|-|-|-|}
\textbf{Taxonomies} & \textbf{Reference} & \textbf{LLM type } & \multicolumn{1}{c|}{\textbf{Pros \& Cons}} \\ 
\hhline{|-|-|-|-|}
\multirow{9}{*}{ \begin{tabular}[c]{@{}c@{}} Extract meaningful \\ features from \\ raw observations \end{tabular}} & \cellcolor{blue!3}\cite{10555960} & \cellcolor{blue!3}{\begin{tabular}[c]{@{}c@{}} LLaVA \\ \cite{NEURIPS2023_6dcf277e} \end{tabular}} & 
\cellcolor{blue!3}\begin{tabular}[c]{@{}l@{}}\textcolor{blue!20}{\ding{108}} LLM-based agents simulate user feedback for personalized network service optimization \\ 
\textcolor{green}{\ding{51}} LLM performs real-time multimodal evaluation to eliminate costly human feedback \\ 
\textcolor{green}{\ding{51}} Adaptable to diverse user preferences and multimodal input scenarios \\ 
\textcolor{green}{\ding{51}} Eliminates reliance on manual reward shaping via LLM-extracted features \\ 
\textcolor{red}{\ding{55}} Multi-agent RL requires per-user LLM feedback, causing overhead as users grow \end{tabular} \\ 
\hhline{|~|-|-|-|}
& \cellcolor{blue!3}\cite{10.1145/3712678.3721880} & \cellcolor{blue!3}{\begin{tabular}[c]{@{}c@{}} GPT-2 \\ T5 \cite{2020t5} \\ Qwen1.5 \\ \cite{bai2023qwen} \end{tabular}} & 
\cellcolor{blue!3}\begin{tabular}[c]{@{}l@{}}\textcolor{blue!20}{\ding{108}} LLM extracts semantic features of bandwidth estimation for DRL's decision-making \\ 
\textcolor{green}{\ding{51}} Leverages LLMs' generalization for unseen states to mitigate online RL interaction risks \\ 
\textcolor{green}{\ding{51}} LLM enhances RL's cross-scenario generalization, improving bandwidth estimation accuracy \\ 
\textcolor{red}{\ding{55}} Dependence on LLM feature quality; misalignment in state encoding can propagate errors \end{tabular} \\ 
\hhline{|~|-|-|-|}
& \cellcolor{blue!3}\cite{lotfi2025oran} & \cellcolor{blue!3}{\begin{tabular}[c]{@{}c@{}} Oransight\\-2.0 \cite{gajjar2025oransight} \\ GPT-2 \end{tabular}} & 
\cellcolor{blue!3}\begin{tabular}[c]{@{}l@{}}\textcolor{blue!20}{\ding{108}} Dual-LLM-augmented MARL for semantic state representation in O-RAN slicing \\ 
\textcolor{green}{\ding{51}} LLM's domain knowledge enables robustness to network state changes for RL \\ 
\textcolor{green}{\ding{51}} Integration of LLMs decreases state dimensionality and improves sample efficiency in RL \\ 
\textcolor{red}{\ding{55}} Dual-LLM inference and RAG retrieval increase latency of RL in edge deployment \end{tabular} \\ 
\hhline{|-|-|-|-|}
\multirow{12}{*}{\begin{tabular}[c]{@{}c@{}} Reduce learning \\ RL complexity by \\ transferring informal \\ natural language \\information into a \\ formal task-specific \\ language \end{tabular}} & \cellcolor{green!3}\cite{10978505} & \cellcolor{green!3}{\begin{tabular}[c]{@{}c@{}} ALBERT \\ \cite{Lan2020ALBERT:} \end{tabular}} & 
\cellcolor{green!3}\begin{tabular}[c]{@{}l@{}}\textcolor{blue!20}{\ding{108}} LLM-driven intent parsing and validation combined with hierarchical RL \\ 
\textcolor{green}{\ding{51}} Enhance HRL scalability by offloading high-level intent interpretation to LLMs \\ 
\textcolor{green}{\ding{51}} Automates translation of natural language intents into structured optimization tasks \\ 
\textcolor{red}{\ding{55}} Ambiguous intents of LLM's response may cause errors in intent extraction \end{tabular} \\ 
\hhline{|~|-|-|-|}
& \cellcolor{green!3}\cite{10876191} & \cellcolor{green!3}{\begin{tabular}[c]{@{}c@{}} ChatGPT \\-3.5 \\ LLaMA \\ 3.1–8B \\ 3.1–70B \end{tabular}} & 
\cellcolor{green!3}\begin{tabular}[c]{@{}l@{}}\textcolor{blue!20}{\ding{108}}Prompt-driven LLMs generate AD actions as state features for the DDQN \\ 
\textcolor{green}{\ding{51}} Achieve faster convergence and higher rewards in RL via LLM-guided exploration \\ 
\textcolor{green}{\ding{51}} Explain decisions via natural language such as collision avoidance rationale \\ 
\textcolor{green}{\ding{51}} Decouple perception and reasoning from RL exploration to reduce sample complexity \\ 
\textcolor{red}{\ding{55}} Larger LLMs (e.g., 70B) incurring high latency are unsuitable for ultra-low-latency V2I \\ 
\textcolor{red}{\ding{55}} Joint decision-making with multiple LLMs has limited scalability to many coupled tasks \end{tabular} \\ 
\hhline{|~|-|-|-|}
& \cellcolor{green!3}\cite{shokrnezhad2025autonomous} & \cellcolor{green!3}{\begin{tabular}[c]{@{}c@{}} LLaMA \\ 3.1–8B \end{tabular}} & 
\cellcolor{green!3}\begin{tabular}[c]{@{}l@{}}\textcolor{blue!20}{\ding{108}} LLM intent parsing and CoT-driven task decomposition for continual RL agent \\ 
\textcolor{green}{\ding{51}} Eliminate manual reward design via LLM-extracted structured tasks \\ 
\textcolor{green}{\ding{51}} Enhance adaptability to dynamic environments with RAG-augmented context \\ 
\textcolor{green}{\ding{51}} Reduce RL exploration complexity through CoT-guided sub-task sequencing \\ 
\textcolor{red}{\ding{55}} Computational overhead from LLM inference and RAG retrieval limits edge deployment \end{tabular} \\ 
\hhline{|-|-|-|-|}
\end{tabular} 
\end{table*}

LLMs trained on large-scale text corpora are able to internalize syntactic, semantic, and factual knowledge through context-aware representation learning, which enables them to process, interpret, and transform complex information structures effectively \cite{10.1145/3664647.3681014}. Meanwhile, LLMs develop an abstract and high-dimensional representation space that captures patterns across diverse modalities and tasks \cite{Li_2022_CVPR}. The representation space supports capabilities such as in-context learning, compositional reasoning, and semantic alignment, allowing LLMs to map unstructured inputs into coherent and structured forms \cite{NEURIPS2022_9d560961}. This grounding allows LLMs to be naturally extendable to RL contexts by serving as a bridge between high-level task descriptions and low-level policy learning.

Specifically, on the one hand, LLMs are capable of extracting meaningful and semantically rich features from raw observational data by leveraging their powerful pre-trained architectures \cite{10834145}. By leveraging their extensive pre-training on diverse datasets, LLMs can automatically identify relevant patterns, abstract high-level semantics, and process unstructured or heterogeneous inputs (such as textual logs, sensor readings, or even symbolic network states),  into structured and task-relevant features \cite{10759588,10614634}. This capability not only reduces the reliance on domain-specific feature engineering but also enhances the agent’s ability to generalize across diverse environments and tasks. For instance, the authors in \cite{10.1145/3712678.3721880} propose ann LLM-enhanced DRL framework for accurate bandwidth estimation (LLM4Band) in real-time communication systems. The LLM4Band leverages the pre-trained semantic representation capabilities of LLMs to extract high-dimensional network state features from raw network flow data (such as packet histories, latency trends and bandwidth). These features are converted into the continuous action for estimating bandwidth. As a result, the LLM-based policy extraction improves the performance of the RL policy network, which achieves a 12.35\% improvement (compared to LSTM baseline \cite{10.1145/3625468.3653068}) in bandwidth estimation accuracy and a 21\% enhancement (compared to Heuristic algorithm \cite{Holmer2016GCC}) in communication quality.

Building on this foundation of feature extraction for RL policy networks, the challenge of sample inefficiency and poor generalization in dynamic O-RAN network slicing is addressed in \cite{lotfi2025oran} through a more sophisticated dual-LLM architecture. The dual-LLM architecture is employed to extract semantically enriched state representations from raw and unstructured network observations such as RF features, QoS targets, and traffic trends. Specifically, a domain-specific LLM pretrained on O-RAN control logs and slicing policies is dynamically used to generate context-aware prompts. These prompts are fused with trainable tokens and processed by a frozen general-purpose LLM (i.e., GPT-2) to output high-level embeddings. These embeddings replace raw observations as inputs for RL agents, enabling the agents to effectively interpret complex wireless environments and concentrate on policy learning. The experimental results show that the proposed scheme achieves approximately 12.5\% higher cumulative reward compared to traditional MARL (i.e., SAC algorithm \cite{pmlr-v80-haarnoja18b}).

Beyond structured network metrics, LLMs also excel at interpreting rich sensory inputs such as images or text in network services. The work in \cite{10555960} addresses the challenge of maximizing subjective Quality of Experience (QoE) in AI-Generated Content. The LLM is used to analyze images and simulate human-like QoE feedback based on the Big Five personality model (i.e., Openness, Conscientiousness, Extraversion, Agreeableness, and Neuroticism) since user preferences for generated images vary significantly based on personality traits. Then, the high-dimensional visual inputs can be converted into scalar reward signals (aesthetic ratings from $0$ to $1$), which serve as rewards for the RL agent to learn on low-dimensional and semantically rich signals. Experimental results show that the LLM-based DRL policy can achieve up to approximately 133\% higher QoE compared to the DRL with a random policy.


On the other hand, LLMs can significantly reduce the learning complexity for RL agents by translating informal natural language inputs in wireless networks, such as communication commands, user requests, or high-level operational guidelines, into structured and task-specific representations \cite{10614634}. This semantic transformation enables RL agents to better interpret and align with high-level objectives that are often conveyed in ambiguous or unstructured human language, which improves the adaptability of RL systems in dynamic or context-sensitive wireless environments \cite{10638533}. The authors in \cite{10978505} employ the LLM as the state perceiver to bridge natural language intents with RL-based network optimization. Specifically, the paper leverages LLMs to parse unstructured natural language intents (e.g., ``optimize throughput by 10\%") and extract structured optimization targets (e.g., throughput, latency, or energy efficiency) and their desired thresholds. This decouples intent interpretation from the RL agent, where the processed intents are then fed into an attention-based hierarchical RL (HRL) framework, which orchestrates network applications to achieve the specified optimization objectives. Simulation results demonstrate that LLMs enable the RL agent to focus on applications most relevant to the operator's intent, which reduces conflicting actions and outperforms traditional HRL (with improvements of 12.02\% in throughput, 26.5\% in delay, and 17.1\% in energy efficiency).

Additionally, the LLM can bridge high-level language commands and semantic feedback in a wirelsee network. The study in \cite{shokrnezhad2025autonomous} integrates LLMs to bridge language commands with RL for resource allocation in autonomous network orchestration. The LLM is employed to translate unstructured inputs (i.e., strategic textual commands and semantic QoE feedback such as minimizing cost or maximizing quality) into formalized task representations. Specifically, based on the chain-of-thought prompt of the LLaMA-3.1-8B model, it maps the command to system-level objectives via step-back questioning, and maps user feedback to quantifiable semantic requirements through similarity scoring to replace conventional bit-level metrics. These processed outputs (e.g., objectives, state histories, and reward-augmented exemplars) form structured inputs for the RL agent to solve the simpler and numeric sub-problem. As a result, the proposed scheme achieves near-optimal resource allocation costs (within 5\% of optimal) in stable conditions.

Expanding further into more complex and real-time decision-making domains, the work in \cite{10876191} showcases how LLMs can simplify joint optimization problems in vehicular networks by translating informal natural language task descriptions into executable actions, where the LLM is leveraged to translate informal natural language task descriptions (e.g., ``maximize speed while minimizing collisions") into formal and executable AD actions (e.g., acceleration, lane changes). Specifically, by integrating in-context learning, the LLM processes structured prompts containing environmental states (e.g., vehicle positions, velocities), predefined action spaces, and curated examples of past experiences (e.g., good/bad decisions). These prompts enable LLMs to derive discrete AD actions for vehicles, which are embedded as part of the state of a Double Deep Q-Network (DDQN) that optimizes V2I base-station selection. The hybrid LLM-DDQN framework achieves 23\% higher AD rewards and 18\% higher V2I rewards compared to conventional DDQN baselines.

\subsection{LLM as Reward Designer} 

\begin{table*}
\centering
\caption{Summary of LLMs as Reward Designers for Enhancing RL \\ CIRCLES (\textcolor{blue!20}{\ding{108}}) DESCRIBE THE METHODS; CORRECT MARKERS (\textcolor{green}{\ding{51}}) AND CROSS MARKERS REPRESENT (\textcolor{red}{\ding{55}}) PROS AND CONS RESPECTIVELY.}
\label{tab:llm-as-reward-designer}
\centering
\begin{tabular}{|c|c|c|l|}
\hhline{|-|-|-|-|}
\textbf{Taxonomies} & \textbf{Reference} & \textbf{LLM type } & \multicolumn{1}{c|}{\textbf{Pros \& Cons}} \\ 
\hhline{|-|-|-|-|}
\multirow{9}{*}{\begin{tabular}[c]{@{}c@{}} Implicit\\ Reward\\ Designer \end{tabular}}
& \cellcolor{blue!3}\cite{10577381} & \cellcolor{blue!3} Specific LLM & 
\cellcolor{blue!3}\begin{tabular}[c]{@{}l@{}}
\textcolor{blue!20}{\ding{108}} LLM generates optimal positions for UAVs, used as a positioning reward signal \\
\textcolor{green}{\ding{51}} Improves network metrics (e.g., +15\% data rate, +20\% connected devices) \\
\textcolor{green}{\ding{51}} Guides agents based on high-level regional characteristics, not just local metrics \\
\textcolor{red}{\ding{55}} Performance is dependent on the LLM's spatial reasoning capabilities \\
\textcolor{red}{\ding{55}} Adds computational overhead from LLM inference during training
\end{tabular} \\ 
\hhline{|~|-|-|-|}

& \cellcolor{blue!3}\cite{10.5555/3692070.3693141} & \cellcolor{blue!3} PaLM 2 XS~~\cite{anil2023palm2technicalreport} & 
\cellcolor{blue!3}\begin{tabular}[c]{@{}l@{}}
\textcolor{blue!20}{\ding{108}} LLM acts as an automated annotator to generate preference data for training a reward model \\
\textcolor{green}{\ding{51}} Highly scalable alternative to expensive and slow human feedback (RLHF) \\
\textcolor{green}{\ding{51}} Achieves performance comparable to RLHF on tasks like summarization \\
\textcolor{green}{\ding{51}} Can improve model harmlessness more effectively than RLHF \\
\textcolor{red}{\ding{55}} Risk of inheriting and amplifying biases from the AI labeler model \\
\textcolor{red}{\ding{55}} Performance is bottlenecked by the quality and alignment of the labeler LLM
\end{tabular} \\ 
\hhline{|-|-|-|-|}

\multirow{8}{*}{\begin{tabular}[c]{@{}c@{}} Explicit\\ Reward\\ Designer \end{tabular}}
& \cellcolor{green!3}\cite{xie2024textreward} & \cellcolor{green!3} GPT-4 & 
\cellcolor{green!3}\begin{tabular}[c]{@{}l@{}}
\textcolor{blue!20}{\ding{108}} LLM generates executable, dense reward code from natural language goals \\
\textcolor{green}{\ding{51}} Produces interpretable and debuggable reward functions (Python code) \\
\textcolor{green}{\ding{51}} Computationally efficient for RL training (no repeated LLM calls) \\
\textcolor{green}{\ding{51}} Performance matches or exceeds human experts on many robotics tasks \\
\textcolor{green}{\ding{51}} Supports iterative human-in-the-loop refinement to fix failure modes \\
\textcolor{red}{\ding{55}} The initial ``first-shot'' generated code may be suboptimal or buggy
\end{tabular} \\ 
\hhline{|~|-|-|-|}

& \cellcolor{green!3}\cite{ma2024eureka} & \cellcolor{green!3} GPT-4 & 
\cellcolor{green!3}\begin{tabular}[c]{@{}l@{}}
\textcolor{blue!20}{\ding{108}} Combines LLM code generation with an evolutionary search algorithm \\
\textcolor{green}{\ding{51}} Autonomously discovers novel and high-performing reward functions \\
\textcolor{green}{\ding{51}} Outperforms human-designed rewards on a majority of complex tasks \\
\textcolor{green}{\ding{51}} Solves tasks previously intractable with manual reward design (e.g., pen-spinning) \\
\textcolor{red}{\ding{55}} Extremely high computational cost due to the evolutionary outer-loop \\
\textcolor{red}{\ding{55}} Requires access to the environment's source code to provide context
\end{tabular} \\ 
\hhline{|-|-|-|-|}
\end{tabular}
\end{table*}

The efficacy of an RL agent is fundamentally determined by the design of its reward function, which is the very signal to defines its purpose and guides its learning process.
Before the advent of LLMs, the design of reward functions in RL was often a manual and domain-specific process, requiring expert knowledge to define appropriate reward signals that align with the desired objectives and constraints of the task~\cite{YAU2012253,9372298,9329087,8303773,10283826,8924617}.
However, LLMs have the potential to revolutionize this process by automating and generalizing reward function design through their advanced natural language understanding and reasoning capabilities.

LLM-based reward designs can be broadly categorized into two main approaches, as shown in Table \ref{tab:llm-as-reward-designer}:
LLMs as an implicit reward designer, where LLMs evaluate or judge the quality of agent actions or policies based on their understanding of the task and context; and LLMs as an explicit reward designer, where LLMs generate structured reward functions based on high-level task descriptions, system constraints, and optimization objectives~\cite{cao2025SurveyLar}.

\subsubsection{Implicit Reward Designer}

For example, \citeauthor{10577381} propose ann LLM-based reward mechanism in multi-agent RL (MARL) for mobile access points organization using unmanned aerial vehicles (UAVs)~\cite{10577381}.
In this work, multiple agents, which represent UAVs relaying data to ground users or UAVs themselves, are trained to relocate their positinons to optimize the average data rate of user devices and the number of connected devices.
To achieve this, the authors propose a Proximal Policy Optimization (PPO)~\cite{10.5555/3600270.3602057} based MARL method, where the reward function consists of two components: the communication reward and the positioning reward.
The communication reward is designed to encourage agents to maximize the above-mentioned hyperparameter-weighted optimization objectives, and the relay intrinsic reward is designed to encourage agents to optimize the relay nodes that serve as bottlenecks in subsequent networks.
Besides, the authors also introduce ann LLM-based positioning reward to guide the agents locate themselves in a optimal position with exhibited regional characteristics.
Specifically, the LLM receives the number of devices within meshed grid cells and the problem description, and generates the positions where the agents should be located.
Combining the communication reward and the LLM-based positioning reward, the proposed method achieves an up to 15\% improvement in average data rate and an up to 20\% improvement in number of connected devices and compared to the baseline method where the LLM is not used.

A prominent paradigm in this category is Reinforcement Learning from AI Feedback (RLAIF)~\cite{10.5555/3692070.3693141}.
RLAIF automates the costly process of Reinforcement Learning from Human Feedback (RLHF) by using a separate, often more powerful, LLM to generate preference labels for text-based tasks such as summarization and dialogue generation instead of human annotators.
A reward model is then trained on this AI-generated data to provide scalable feedback for policy optimization.

This approach can achieve performance comparable to RLHF on tasks such as text summarization and dialogue generation.
For example, RLAIF and RLHF are preferred over traditional supervised fine-tuning methods with 71\% and 73\% of the win rates for summarization, and 63\% and 64\% for dialogue generation, respectively.
Moreover, RLAIF achieves a higher harmlessness rate of 88\% compared to 76\% for RLHF in dialogue generation tasks, thereby helping to address the scalability limitations of human annotation.

\subsubsection{Explicit Reward Designer}

In contrast to implicit methods, an explicit reward designer uses an LLM to generate an entire, executable reward function, typically as a Python program. This approach offers significant advantages in terms of interpretability, as the reward logic is transparent and can be inspected or debugged by a human. Furthermore, it is computationally efficient during RL training, as it eliminates the need for repeated, costly API calls to an LLM.

A foundational framework in this area is Text2Reward~\cite{xie2024textreward}, which automates the generation of dense, shaped reward functions from a natural language goal and a compact description of the environment's API.
Text2Reward has demonstrated strong performance on various robotics manipulation and locomotion benchmarks, with its generated rewards often matching or even surpassing those written by human experts.

{

For instance, in manipulation tasks, Text2Reward's generated rewards achieved similar or better performance than human-coded rewards in 13 out of 17 tasks, and in locomotion tasks, which include six motor control tasks
including lifting a cube, picking a cube, turning a faucet, pushing a chair, opening a cabinet door, and opening a cabinet darwer,
it learned six locomotion actions with success rates of 94\% or higher.
The framework also supports iterative refinement, allowing a human to provide feedback on failure modes to improve the generated reward code.

}
Building on this, the EUREKA framework~\cite{ma2024eureka} introduces a higher level of autonomy by combining LLM code generation with an evolutionary search algorithm.
EUREKA operates in a loop: it first uses an LLM to generate a population of candidate reward functions, often by providing the environment's source code as context.
These functions are then evaluated by training RL agents in parallel.
Finally, an automated ``reward reflection'' step summarizes the training outcomes and feeds this summary back to the LLM, which then generates an improved set of reward functions for the next iteration.
This meta-learning approach has successfully designed rewards for highly complex dexterity tasks, such as simulated pen-spinning, that were previously intractable with manual reward engineering, where it ouperformed human-designed rewards in 24 out of 29 tasks.

\subsection{LLM as Decision-maker} 

\begin{table*}[]
\centering

\caption{Summary of LLMs as Decision Makers for Enhancing RL \\ CIRCLES (\textcolor{blue!20}{\ding{108}}) DESCRIBE THE METHODS; CORRECT MARKERS (\textcolor{green}{\ding{51}}) AND CROSS MARKERS (\textcolor{red}{\ding{55}}) REPRESENT PROS AND CONS RESPECTIVELY.}
\label{tab:decisionmaker}
\centering
\begin{tabular}{|p{2cm}|c|c|l|}
\hhline{|-|-|-|-|}
\textbf{Taxonomies} & \textbf{Reference} & \textbf{LLM type } & \multicolumn{1}{c|}{\textbf{Pros \& Cons}} \\ 
\hhline{|-|-|-|-|}
\multirow{12}{*}{\begin{tabular}[c]{@{}c@{}} LLM-based \end{tabular}}
& \cellcolor{blue!3}\cite{han2025agent} & \cellcolor{blue!3}{\begin{tabular}[c]{@{}c@{}} GPT-4o-mini \end{tabular}}  & 
\cellcolor{blue!3}\begin{tabular}[c]{@{}l@{}}\textcolor{blue!20}{\ding{108}} Multi-agent hierarchical framework (ACMA) for adaptive HAPS fleet positioning \\ 
\textcolor{green}{\ding{51}} Decomposes complex optimization into interpretable subtasks for improved scalability \\ 
\textcolor{green}{\ding{51}} Can interpret high-level natural language instructions into efficient, actionable strategies \\  
\textcolor{red}{\ding{55}} Increased system complexity due to multiple specialized agents and inter-agent coordination \\ 
\textcolor{red}{\ding{55}} Potential response latency overhead from orchestrating multiple LLMs per decision cycle \end{tabular} \\ 
\hhline{|~|-|-|-|}
& \cellcolor{blue!3}\cite{emami2025llm} & \cellcolor{blue!3}{\begin{tabular}[c]{@{}c@{}} GPT-o3-mini \\ GPT-4o-mini \\ GPT-2.5-Trubo \end{tabular}}    & 
\cellcolor{blue!3}\begin{tabular}[c]{@{}l@{}}\textcolor{blue!20}{\ding{108}} ICL-based Data Collection Scheduling (ICLDC) for UAV-assisted sensor networks \\ 
\textcolor{green}{\ding{51}} Implements a feedback loop by incorporating real execution results into in-context prompts \\ 
\textcolor{green}{\ding{51}} Adaptation occurs via prompt update rather than costly weight updates or retraining \\ 
\textcolor{red}{\ding{55}} Vulnerable to prompt-based attacks (e.g., jailbreaking) which can degrade network performance \\ 
\textcolor{red}{\ding{55}} Effectiveness depends on the feedback window, may be challenging with limited context length
 \end{tabular} \\ 
\hhline{|~|-|-|-|}
& \cellcolor{blue!3}\cite{10976336} & \cellcolor{blue!3}{\begin{tabular}[c]{@{}c@{}} ViT \cite{dosovitskiy2020image} \\\end{tabular}}  & 
\cellcolor{blue!3}\begin{tabular}[c]{@{}l@{}}\textcolor{blue!20}{\ding{108}} Multi-modal LLM-powered joint decision-making for ramp merging in vehicular networks \\ 
\textcolor{green}{\ding{51}} Multi-modal encoder captures both textual and visual context, improving situational awareness \\ 
\textcolor{green}{\ding{51}} Enforced reasoning and explicit meta-decision generation foster transparency and interpretability \\ 
\textcolor{red}{\ding{55}} Fine-tuning and multi-agent rollout still require significant data engineering and simulation effort \\ 
\textcolor{red}{\ding{55}} Real-time deployment may be challenged by large LLMs and vision processing on edge hardware \end{tabular} \\ 
\hhline{|-|-|-|-|}

\multirow{13}{*}{\begin{tabular}[c]{@{}c@{}} LLM-guiding \end{tabular}}
& \cellcolor{green!3}\cite{shokrnezhad2025autonomous} & \cellcolor{green!3}{\begin{tabular}[c]{@{}c@{}} LLaMA \\ 3.1–8B  \end{tabular}}  & 
\cellcolor{green!3}\begin{tabular}[c]{@{}l@{}}\textcolor{blue!20}{\ding{108}} Hierarchical LLM-RL framework for resource allocation in SAGIN \\ 
\textcolor{green}{\ding{51}} Combines high-level LLM strategy generation with efficient low-level RL execution  \\ 
\textcolor{green}{\ding{51}} Employs CoT reasoning and few-shot learning to improve reliability and mitigate hallucinations \\ 
\textcolor{red}{\ding{55}} Added complexity from multi-layer (LLM+RL) architecture and prompt engineering requirements \end{tabular} \\ 
\hhline{|~|-|-|-|}
& \cellcolor{green!3}\cite{xu2025scalable} & \cellcolor{green!3}{\begin{tabular}[c]{@{}c@{}}Not Specified \end{tabular}} & 
\cellcolor{green!3}\begin{tabular}[c]{@{}l@{}}\textcolor{blue!20}{\ding{108}} LLM-guided knowledge distillation framework for efficient RL training in UAV networks
 \\ 
\textcolor{green}{\ding{51}} Utilizes LLM's high-level decision knowledge and CoT reasoning to accelerate RL training \\ 
\textcolor{green}{\ding{51}} Reduces inefficient RL exploration and risk of local optimum convergence \\ 
\textcolor{red}{\ding{55}} Effectiveness depends on the quality and generalizability of LLM-generated teaching data \\ 
\textcolor{red}{\ding{55}} Policy optimization from LLM outputs may suffer from biased distributions due to hallucinations \end{tabular} \\ 
\hhline{|~|-|-|-|}
 & \cellcolor{green!3}\cite{10884745} & \cellcolor{green!3}{\begin{tabular}[c]{@{}c@{}} LLaMA  \end{tabular}} & 
\cellcolor{green!3}\begin{tabular}[c]{@{}l@{}}\textcolor{blue!20}{\ding{108}} PPAE algorithm that integrates LLaMA into RL actor network training for AUV path planning \\ 
\textcolor{green}{\ding{51}} Action probability distribution generated by LLaMA narrows the exploration space in training \\ 
\textcolor{green}{\ding{51}} Effective distributional priors reduce policy variance in complex marine environments \\ 
\textcolor{red}{\ding{55}} Policy optimization from LLM outputs may suffer from biased distributions due to hallucinations \end{tabular} \\ 
\hhline{|~|-|-|-|}
  & \cellcolor{green!3}\cite{10819462} & \cellcolor{green!3}{\begin{tabular}[c]{@{}c@{}} Gemma \\ 2-2B  \end{tabular}} & 
\cellcolor{green!3}\begin{tabular}[c]{@{}l@{}}\textcolor{blue!20}{\ding{108}} Lightweight LLM-assisted DQN framework for task scheduling in multi-cloud environments  \\ 
\textcolor{green}{\ding{51}} Generate action candidates to narrow exploration space and accelerate convergence  \\ 
\textcolor{green}{\ding{51}} Lightweight LLM (2B parameters) enables deployment on resource-constrained devices  \\ 
\textcolor{red}{\ding{55}} Low-parameter LLMs may be limited in knowledge scope when processing complex state \end{tabular} \\ 
\hhline{|-|-|-|-|}
\end{tabular}


\end{table*}

Benifit from the scaling law~\cite{kaplan2020scaling}, LLMs demonstrate strong understanding capabilities on out-of-distribution (OOD) data, enabling them to make decisions in complex and rare environments~\cite{li2022pre}. Furthermore, with the emergence of several key capabilities, including multimodal understanding~\cite{wu2024next}, task planning~\cite{kannan2024smart}, and sequential reasoning~\cite{wei2022chain}, LLMs are expected to enhance or even substitute traditional RL as a better decision-maker~\cite{li2022pre, zhou2024large} rather than as a standalone tools. In this section, the reviewed methods are categorized into two main types based on the role of LLMs in decision-making: 1) LLM-based methods that LLM directly generate actions and 2) LLM-guiding methods that LLM guides the RL agent to make decisions.

\subsubsection{LLM-based}
Early studies have recognized the powerful reasoning and decision-making capabilities of LLMs, and have successfully embedded them into various types of network optimization and resource allocation tasks~\cite{han2025agent, emami2025llm, shokrnezhad2025autonomous, xu2025scalable}. Compared with traditional RL-based methods, these approaches have achieved significant improvements in key metrics such as QoS, coverage, and throughput.

~\citeauthor{han2025agent}~\cite{han2025agent} leverages LLMs to operate a fleet of High Altitude Platform Stations (HAPS)~\cite{widiawan2007high} for dynamic user coverage in wireless networks. The proposed framework, called Action Coordination and Multi-Agent (ACMA), utilizes LLMs to optimize the positions of HAPS airships to provide a more efficient and relaible service.
ACMA decomposes the complex optimization problem into a series of four sub-tasks, each handled by a specialized LLM-based agent~\cite{zhao2024expel}. The workflow involves a Data Analysis Agent that identifies high-demand areas, a Target Location Selection Agent that makes the primary positioning decisions, an Overlap Avoidance Agent that refines these positions to enhance resource efficiency, and a Special Event Handling Agent. Furthermore, the ACMA framework supports to plans and adjusts the stations like an autonomous agent, based on occasional high-level natural language instructions.
This hierarchical approach allows for more nuanced decision-making, enabling the system to adapt to dynamic user demands and environmental changes effectively. ACMA demonstrates a 130\% improvement in user coverage compared to RL-based method and random walk~\cite{lawler2010random}.
Similarly, ~\citeauthor{emami2025llm}~\cite{emami2025llm} introduce an In-Context Learning-based Data Collection Scheduling (ICLDC) scheme for UAV-assisted sensor networks~\cite{gao2023aoi}, offering an alternative to DRL methods in time-sensitive emergency scenarios. In this framework, the UAV collects real-time network data and relies on an edge-hosted LLM, which, guided by a dynamic context of task descriptions and in-context examples, directly generates scheduling decisions. ICLDC introduces a continuous feedback mechanism: after each scheduling decision, the outcome is incorporated back into the LLM’s context, enabling the model to iteratively adapt its decisions based on recent performance—effectively achieving online adaptation without weight updates. This approach allows the LLM to adjust its actions responsively in non-stationary environments, bridging the gap between static pre-trained models and the demands of real-time edge intelligence. Compared with traditional methods that rely solely on channel conditions, ICLDC achieves a significant reduction (up to 56\%) in packet loss. The study also highlights LLMs' potential vulnerability to adversarial prompt manipulation, suggesting that developing robust defense strategies is an important avenue for future research.

In addition to scheduling in aerial networks, LLMs have been explored for decision making in vehicular networks. For instance, ~\citeauthor{10976336}~\cite{10976336} propose AgentsCoMerge, a framework for ramp merging tasks that leverages LLMs equipped with a multi-modal encoder. By incorporating visual as well as textual information, AgentsCoMerge enables the LLM to better interpret vehicle surroundings, enhancing its ability to make spatially and contextually informed decisions. The framework enforces explicit reasoning for decision transparency and reliability, requiring the LLM to justify high-level meta-decisions, which are then converted by each agent into executable trajectories. In cases of failure (e.g., collisions), the LLM receives feedback and reasoning prompts, iteratively refining its decisions based on detailed analysis of mistakes. To better align with domain-specific requirements, AgentsCoMerge also employs fine-tuning using LoRA~\cite{hu2022lora}, improving the LLM's ramp merging expertise with minimal computational overhead. This approach highlights the potential of LLMs as adaptive and explainable decision makers in complex, real-time vehicular network environments.

{

\textbf{2) LLM-guiding}:
Although these LLM-based methods demonstrate the potential of LLMs as decision-makers, they still face several challenges since almost all resource schedule and allocation are dictatorially determined by LLMs rather than RL agents. However, the computational-intensive nature of LLM makes it unable to respond promptly to the fast-changing wireless channel resource environment, resulting in its inability to handle most communication resource allocation tasks with low latency requirements~\cite{jiang2018low}. Furthermore, with the issue of LLM hallucinations, the reliability of its responses is gradually being questioned~\cite{liu2024exploring}, especially when applied to an OOD scenario~\cite{yao2023llm}. Therefore, several recent studies have attempted to leverage LLMs to guide the decision-making process of RL agents, rather than replacing them entirely.

For example, ~\citeauthor{shokrnezhad2025autonomous}~\cite{shokrnezhad2025autonomous} introduces a framework called Autonomous Reinforcement Coordination (ARC) for orchestrating a Space-Air-Ground Integrated Networks (SAGIN). ARC decomposees the orchestration task into two layers: a high-level LLM-based layer and a low-level RL-based layer. The LLM layer generates high-level strategies and guidance as the prior information, while the RL layer executes actions based on these strategies.
For instance, the LLM first generate a user sequence that includes user ID, service ID and its action profile, then the RL agent will execute these action selection for each user.
This hierarchical approach allows for more efficient and adaptive resource allocation in dynamic environments. Besides, ARC emploies Chain-of-Thought (CoT) reasoning~\cite{wei2022chain} through few-shot learning~\cite{song2023llm} to reduce the risk of hallucinations to enhance system robustness and reliability. Specifically, a few reasoning exemplars are demonstrated in LLM prompts to provocide a more detailed instruction, which includes a chain of thoughts demonstrating the sequence of users and their corresponding actions selected for a specific history-objective pair. Compared with Non-Reinforcement ARC that disables the RL layer and allows the LLM to directly control the low-level allocations, the proposed ARC framwork demonstrates a significant less fluctuations brought by LLN hallucinations. Therefore, ARC strikes a balance between the high-level decision-making capabilities of LLMs and the low-level execution efficiency of RL agents, achieving a more robust and reliable orchestration in SAGIN.

Compared to involving LLMs during the inference phase, \citeauthor{xu2025scalable}~\cite{xu2025scalable} attempted to leverage LLMs to optimize the RL training process in an UAV-enabled multi-hop networking. The proposed LLM-guided knowledge distillation framework shares a similar intuition that guides the RL agent to learn the optimal UAV scheduling strategy based on the high-level decision knowledge from LLM.} However, the knowleage transfer process is conducted during the training phase of RL agents, where a knowledge distillation process is designed to prepare decisions that align with common sense for RL agents. Specifically, the LLM acts as a teacher model, reasoning out appropriate actions based on a carefully designed Chain-of-Thought (CoT) process and the current state. These state-action pairs are collected then used to teach the RL agent to perform decision. This approach effectively avoids the issue of the RL agent engaging in a large amount of inefficient exploration in the early stages of training under random initialization, while also reducing the risk of the RL agent prematurely converging to a local optimum. In a 3.5 Km $\times$ 3.5Km simulation environment with 18 UAVs and 150 UEs, the proposed framework achieves around an average of 23\% higher UE coverage, 52\% higher data rate, and a 19\% higher UAV availability ratio compared to three multi-agent reinforcement learning (MARL) baselines, GVis~\cite{zhang2022cooperative}, GA2C~\cite{li2022deep} and MAPPO~\cite{yu2022surprising}, that without the LLM guidance.

In addition to serving as a teacher model to reason about the RL agent's actions \cite{xu2025scalable}, an LLM can also act as a soft expert policy to guide RL training. The authors in \cite{10884745} integrate the LLaMA model into the RL pipeline to enhance decision robustness in autonomous underwater vehicle (AUV) path planning. Specifically, they design a proximal policy advantage estimation (PPAE) algorithm, where the LLaMA model is utilized to generate the action probability distribution based on environmental inputs. The action probability functions as the soft expert policy to narrow the exploration space of RL and guide the actor network training to benefit from the LLM’s prior knowledge. Thus, the RL agent can optimize the actor’s policy distribution based on both the guidance from the LLaMA model and the original reward signals. Compared with the traditional PPO algorithm, the proposed PPAE improves episode rewards by approximately 11.1\% to 27.3\% and achieves shorter path planning, with path lengths reduced by about 16.7\% to 20\%.

Apart from leveraging LLMs to generate the action probability distribution for RL agents, LLMs can also narrow down the exploration space by producing high-quality action candidates in complex system optimization. The work in \cite{10819462} proposes a lightweight LLM-assisted DQN framework for task scheduling in multi-cloud environments. In this framework, the LLM acts as a decision guide, with its core function being the generation of an action candidate set to enhance the decision-making efficiency of the RL agent. Specifically, during the RL training process, the fine-tuned LLM receives the current system state—including the task queue, available resources, and optimization objectives—and generates a high-quality action candidate set (i.e., a group of recommended schedulable instances). The DQN agent then performs Q-value evaluation and makes final decisions within this refined, high-quality candidate set, thereby avoiding inefficient exploration in the vast original action space. This approach effectively combines the LLM's capabilities in semantic reasoning and domain knowledge with DQN's strengths in numerical optimization and long-term return trade-offs. Experimental results demonstrate that the proposed LLM-guided framework significantly outperforms baseline methods \cite{XIA202238} in terms of cost, makespan, and energy consumption.

In addition, the recent study in \cite{10839306} has shown that LLMs also exhibit great potential in solving multi-objective optimization problems, where their ability to act as planners can be leveraged to assist RL in policy learning and action selection. Specifically, the work in \cite{10839306} formulates a multi-objective optimization problem (MOP) that jointly optimizes UAV deployment and transmission power control, which is decomposed into a series of optimization sub-problems. To simultaneously work on these sub-problems, an LLM is integrated as a closed-box search operator and guided by MOP-specific prompts that consider the objective functions and constraint satisfaction to generate Pareto-optimal solutions. Experiments show that convergence speed improves by approximately 2.7\% compared with the baseline reference vector guided evolutionary algorithm \cite{9419072}.

\subsection{LLM as Generator} 

\begin{table*}[]
\centering

\caption{Summary of LLMs as Generator for Enhancing RL \\ CIRCLES (\textcolor{blue!20}{\ding{108}}) DESCRIBE THE METHODS; CORRECT MARKERS (\textcolor{green}{\ding{51}}) AND CROSS MARKERS (\textcolor{red}{\ding{55}}) REPRESENT PROS AND CONS RESPECTIVELY.}
\label{tab:generator}
\centering
\begin{tabular}{|p{2cm}|c|c|l|}
\hhline{|-|-|-|-|}
\textbf{Taxonomies} & \textbf{Reference} & \textbf{LLM type } & \multicolumn{1}{c|}{\textbf{Pros \& Cons}} \\ 
\hhline{|-|-|-|-|}
\multirow{9}{*}{\begin{tabular}[c]{@{}c@{}} Word Model \end{tabular}}
& \cellcolor{blue!3}\cite{dharmalingam2025aero} & \cellcolor{blue!3}{\begin{tabular}[c]{@{}c@{}} OPT \\ LLaMA-2 \\ TimeNet \\ Time-LLM \end{tabular}}  & 
\cellcolor{blue!3}\begin{tabular}[c]{@{}l@{}}\textcolor{blue!20}{\ding{108}} Integrate multiple LLMs to enhance UAV mission security and operational efficiency \\ 
\textcolor{green}{\ding{51}} Adopt diverse smaller mission-critical LLMs to reduce latency and improve real-time decisions \\ 
\textcolor{green}{\ding{51}} Distributed architecture offloads computational tasks across on-board, edge, and cloud \\ 
\textcolor{red}{\ding{55}} Each domain-specific LLM requires its own training and fine-tuning, increasing complexity
 \end{tabular} \\ 
\hhline{|~|-|-|-|}
& \cellcolor{blue!3}\cite{mohsin2025retrieval} & \cellcolor{blue!3}{\begin{tabular}[c]{@{}c@{}} GPT-4o \\ GPT-4.0 \\ Gemini-\\1.5-flash \end{tabular}}    & 
\cellcolor{blue!3}\begin{tabular}[c]{@{}l@{}}\textcolor{blue!20}{\ding{108}} A pre-processing pipeline for wireless environment perception through RAG-based LLM \\ 
\textcolor{green}{\ding{51}} Utilizes RAG to enhance LLMs' understanding of wireless networks from multiple data sources \\ 
\textcolor{red}{\ding{55}} Performance highly dependent on the performances of the RAG model and LLMs
 \end{tabular} \\ 
\hhline{|~|-|-|-|}
& \cellcolor{blue!3}\cite{yang2025llmkey} & \cellcolor{blue!3}{\begin{tabular}[c]{@{}c@{}} GPT-4 \\\end{tabular}}  & 
\cellcolor{blue!3}\begin{tabular}[c]{@{}l@{}}\textcolor{blue!20}{\ding{108}} An LLM-driven framework for generating cryptographic keys in IoV systems. \\ 
\textcolor{green}{\ding{51}} Improve key generation efficiency while maintaining crucial channel information
\end{tabular} \\ 
\hhline{|~|-|-|-|}
& \cellcolor{blue!3}\cite{shao2024wirelessllm} & \cellcolor{blue!3}{\begin{tabular}[c]{@{}c@{}} Claude-3 Opus \\GPT-4 \\\end{tabular}}  & 
\cellcolor{blue!3}\begin{tabular}[c]{@{}l@{}}\textcolor{blue!20}{\ding{108}} A framework for adapting and enhancing LLMs in the wireless communication networks \\ 
\textcolor{green}{\ding{51}} Flexible enhancement approaches that only require closed-box access of LLMs \\ 
\textcolor{red}{\ding{55}} The security and privacy are not guaranteed as the LLM may leak sensitive information \\ 
\textcolor{red}{\ding{55}} The computational overhead raises concerns while deploying on diverse edge devices
\end{tabular} \\ 
\hhline{|-|-|-|-|}

\multirow{1}{*}{\begin{tabular}[c]{@{}c@{}} Policy Interpreter \end{tabular}}
& \cellcolor{green!3}\cite{ali2023leveraging} & \cellcolor{green!3}{\begin{tabular}[c]{@{}c@{}} Falcon-7B  \end{tabular}}  & 
\cellcolor{green!3}\begin{tabular}[c]{@{}l@{}}\textcolor{blue!20}{\ding{108}} Integrate LLMs into ZTNs to enhance the transparency of RL and improve user interactions \\ 
\textcolor{green}{\ding{51}} Utilize LLMs to transform perplexing network actions into digestible insight \\ 
\textcolor{red}{\ding{55}} The maintenance and retrieval of massive data reduce the practicality of real-time scenarios   \\ 
\textcolor{red}{\ding{55}} Using LLMs to process massive sensitive data will introduce privacy and ethical concerns \end{tabular} \\ 
\hhline{|-|-|-|-|}
\end{tabular}


\end{table*}



{

LLMs can serve as generators in various ways, such as generating synthetic training data, simulating environments, providing high-level abstractions, or formulating explanation that facilitate RL learning. In this section, we categorize we categorize the reviewed methods into two main types based on the function of LLMs: 1) LLM as a world model simulator, where LLMs simulate the environment dynamics or generate synthetic data; and 2) LLM as a policy interpreter, where LLMs interpret high-level policies or provide explanations to guide RL agents.

\subsubsection{ Word Model Simulator}
The application of RL to real-world wireless networks is fundamentally constrained by the practicalities of data collection. Training an RL agent on a live network is often prohibitively expensive, time-consuming, and carries significant operational risks~\cite{deng2022dreamerpro, matsuo2022deep}, such as service disruption or instability. Simulation offers a safe, scalable, and efficient alternative, allowing agents to undergo millions of training episodes in a controlled virtual environment. However, the efficacy of a simulation-trained agent is contingent upon the fidelity and diversity of the simulation. A significant gap between the simulated environment and real-world dynamics—often termed the ``sim-to-real" gap—can lead to policies that perform poorly upon deployment.

Existing studies have attempted to adopt machine learning (ML) models to model and synthesize wireless data~\cite{orekondy2022mimo, he2025attention, pandey2023resource}, such as ~\citeauthor{orekondy2022mimo}~\cite{orekondy2022mimo} adopt Generative Adversarial Networks (GANs)~\cite{goodfellow2014generative} to generate channel impulse response. However, prior ML models often lack a deeper contextual understanding, which can result in unrealistic or overly simplistic simulations that fail to capture the complexities of real-world wireless environments. In contrast, LLMs can utilize their extensive world knowledge to implicitly infer causal relationships between prompt concepts and expected data characteristics, thereby generating data from high-level semantic prompts.
For example, ~\citeauthor{dharmalingam2025aero}~\cite{dharmalingam2025aero} propose a distributed system engineered to improve the security and operational efficacy of UAV during mission-critical operations. This system adopts a team of diverse, smaller LLMs, each fine-tuned for a specific generative task, including real-time data inference, anomaly detection, and activities forecasting. Moreover, these ``special-skilled'' LLMs employ a combination of Supervised Fine-Tuning (SFT)~\cite{prottasha2022transfer} and Reinforcement Learning from Human Feedback (RLHF)~\cite{ouyang2022training}. This hybrid approach ensures that the LLMs are well-adapted to the specific data and operational context of UAV missions. The generated data can be used to train task-specific RL agents, enabling them to learn effective policies in a simulated environment that closely mirrors real-world dynamics.

~\citeauthor{mohsin2025retrieval}~\cite{mohsin2025retrieval} develop a wireless environment perception framework that translate raw, multi-modal sensor data into a structured, human-readable, and machine-actionable textual description. This perception is positioned as an essential input for higher-level RL tasks, such as global optimization, resource allocation, and intelligent handover decisions, which require a comprehensive understanding of the operational context. A Retrieval-Augmented Generation (RAG) architecture~\cite{lewis2020retrieval} is employed in this system, which ingests a diverse set of synchronized data streams from sensors commonly found in autonomous vehicle or smart city environments, including high-resolution cameras, GPS units, and LiDAR scanners. By processing and fusing this information, the LLM aims to generate a detailed narrative of the environment, identifying objects, their spatial relationships, and potential impacts on wireless signal propagation. Compared to a ``Vanilla LLM" baseline that without the RAG-retrieved context, the proposed framework demonstrates an 8\% improvement in relevancy, an 8\% improvement in faithfulness (adherence to source facts), and a 12\% improvement in overall accuracy.

Beyond applying LLMs for generating or integrating wireless channel data, emerging research has begun exploring the potential of leveraging LLMs for wireless key generation~\cite{yang2025llmkey}, which is essential for extablishing secure communication links in Next-Gen Internet of Vehicles (IoV) systems~\cite{zhang2016key}. Currenty key generation methods typically require excessive and repeated channel probing, leading to significant overhead and latency~\cite{wang2011fast}.
LLMKey is designed to address this challenge by skipping channel probing and leveraging the LLM's few-shot ability to generate values for skipped channel probing.
The experimental results demonstrate that even with a probing ratio as low as $0.5$, the system still attains an average key agreeement rate of $0.978$. However, the LLM-driven key generation process is still limited by the high computational overhead and latency of LLMs, which may not be suitable for real-time applications. Therefore, adopting knowledage distillation techniques~\cite{gou2021knowledge} to transfer the LLM's knowledge to a lightweight RL agent is a more promising solution to address this issue.


However, existing methods are still limited by the LLM's inherent capabilities, which may not be sufficient to fully understand the complex and dynamic nature of wireless environments. Therefore, ~\citeauthor{shao2024wirelessllm}~\cite{shao2024wirelessllm} release WirelessLLM, a comprehensive framework to further enhance the domain-specific knowleage and expertise of LLMs in wireless systems by incorporated several advanced techniques. WirelessLLM first involves carefully crafting CoT prompts and few-shot task demonstrations to guide the LLM's reasoning process. Then, WirelessLLM allows LLMs to leverage external software tools, such as MATLAB or Python functions to perform complex calculations or simulations that are beyond the LLM's inherent capabilities.
The generative capabilities of LLMs present a novel opportunity to create richer, more realistic, and more strategically complex simulations, thereby narrowing this gap and enhancing the robustness of RL agents.

\subsubsection{ Policy Interpreter}
Despite the performance benefits of RL, its adoption in critical infrastructure like wireless networks is hindered by a significant trust deficit. RL agents often operate as ``closed-boxes", with their decision-making rooted in complex, high-dimensional neural networks that are opaque to human operators~\cite{iyer2018transparency}. This lack of transparency is a major barrier to deployment, as network engineers and operators are justifiably reluctant to cede control of mission-critical systems to AI whose reasoning cannot be understood, verified, debugged, or trusted~\cite{otto2023deep}. Explainable Reinforcement Learning (XRL) has been introduced to address this challenge, and LLMs, with their innate mastery of human language, can be used for generating policy explanations~\cite{puiutta2020explainable}. For instance, ~\citeauthor{lin2024learning}~\cite{lin2024learning} construct a decision tree based on the policy, and then use LLMs to derive explanation based on the decision three. ~\citeauthor{das2023state2explanation}~\cite{das2023state2explanation} introduce State2Explanation, a framework that leverages a specific model to generate concept-based explanations based on the state-action pairs.
Although several methods have been proposed to enhance the interpretability of RL policies, only limited research has deployed this paradigm in the scenario of wireless networks. A canonical example is presented by ~\citeauthor{ali2023leveraging} in their work on DRL-based anti-jamming strategies for Zero Touch Networks~\cite{ali2023leveraging}. In their framework, a DRL agent learns a policy to switch communication channels to evade a jammer. To make the agent's decisions transparent, the system feeds the state (the vector of received jamming power across channels), the agent's chosen action (the selected channel), and the resulting reward to an LLM. The LLM is then prompted to generate a textual explanation for that specific decision.
}

\subsection{Lesson Learned.} The integration of LLMs into RL establishes a new paradigm that elevates wireless network optimization from experience-driven learning to knowledge-guided reasoning. By functioning as state perceiver, reward designer, decision-maker, and generator, LLMs enrich RL with semantic understanding, adaptive objective formulation, and enhanced environmental modeling. However, these benefits must be balanced against practical challenges. Hallucination may introduce inconsistent or biased outputs, necessitating safeguards such as structured prompting, schema-constrained generation, multi-sample validation, and stability checks to prevent cumulative errors. Moreover, LLMs of different sizes may exhibit varying effectiveness across the four roles, suggesting that lightweight or distilled models can be used for perception and reward shaping, whereas reasoning-intensive tasks may require larger models. Finally, the latency of LLM inference must be reconciled with the ultra-low-latency demands of wireless networks, motivating hybrid designs where LLMs operate offline or asynchronously while real-time control relies on compact distilled policies. Overall, these considerations highlight the multiple trade-offs of LLM-enhanced RL for future wireless networks.

\section{LLM-enhanced RL for Low-altitude economy networking}

\subsection{Background and Motivation}

Low-altitude economy networking (LAENet) refers to the next-generation network paradigm that integrates communication and computation infrastructure to support the operation of manned and unmanned aerial vehicles (UAVs) within airspace below 1,000 meters \cite{cai2025secure,11024060}. The primary vision of LAENet is to unlock economy and societal value through the flexible deployment of aerial platforms across various verticals, including intelligent transportation systems, emergency services, and ubiquitous wireless coverage \cite{11073138}. 
Despite these advantages, the LAENet faces several complex challenges. First, the need for real-time decision-making in UAV control and multi-agent coordination requires high responsiveness under dynamic conditions \cite{10599389}. Second, the wireless environment is subject to rapid changes due to user mobility, terrain variations, and fluctuating channel conditions, making it difficult to maintain consistent connectivity \cite{Ahmad2025}. In addition, the LAENet often operates under constraints related to energy, bandwidth, and computational power for resource-limited UAVs and edge devices \cite{JIN2023100594}.


To tackle these complex challenges, many existing studies have shown that RL offers a promising approach for LAENet, enabling autonomous, model-free, and adaptive control. By interacting with dynamic environments, RL can learn robust and time-sensitive decision policies to efficiently manage limited energy, bandwidth, and computational resources \cite{10283826,cai2025secure}. Nevertheless, classical RL still exhibits some limitations in practical LAENet deployments. First, the limited generalization ability often fails to adapt to unseen tasks and environments in the LAENet since the models are trained in specific settings \cite{NEURIPS2020_5a751d6a,10143407}. Second, reward function design relies on manual specification, where some poorly designed signals can lead to inefficient behaviors and misguide learning for RL agent \cite{NEURIPS2022_266c0f19,cai2025secure}. In addition, DRL suffers from stability issues due to model approximation errors and lacks transparency in decision-making \cite{moerland2023model,HE2021107052}, posing significant risks for safety-critical LAENet applications. Recent advancements in LLMs offer a compelling opportunity to augment DRL in addressing these limitations. Then, we provide a case study to show the use of LLM-enhanced RL for the LAENet optimization in the next part.

\begin{figure}[t]
\centering
\includegraphics[width=1.01\linewidth]{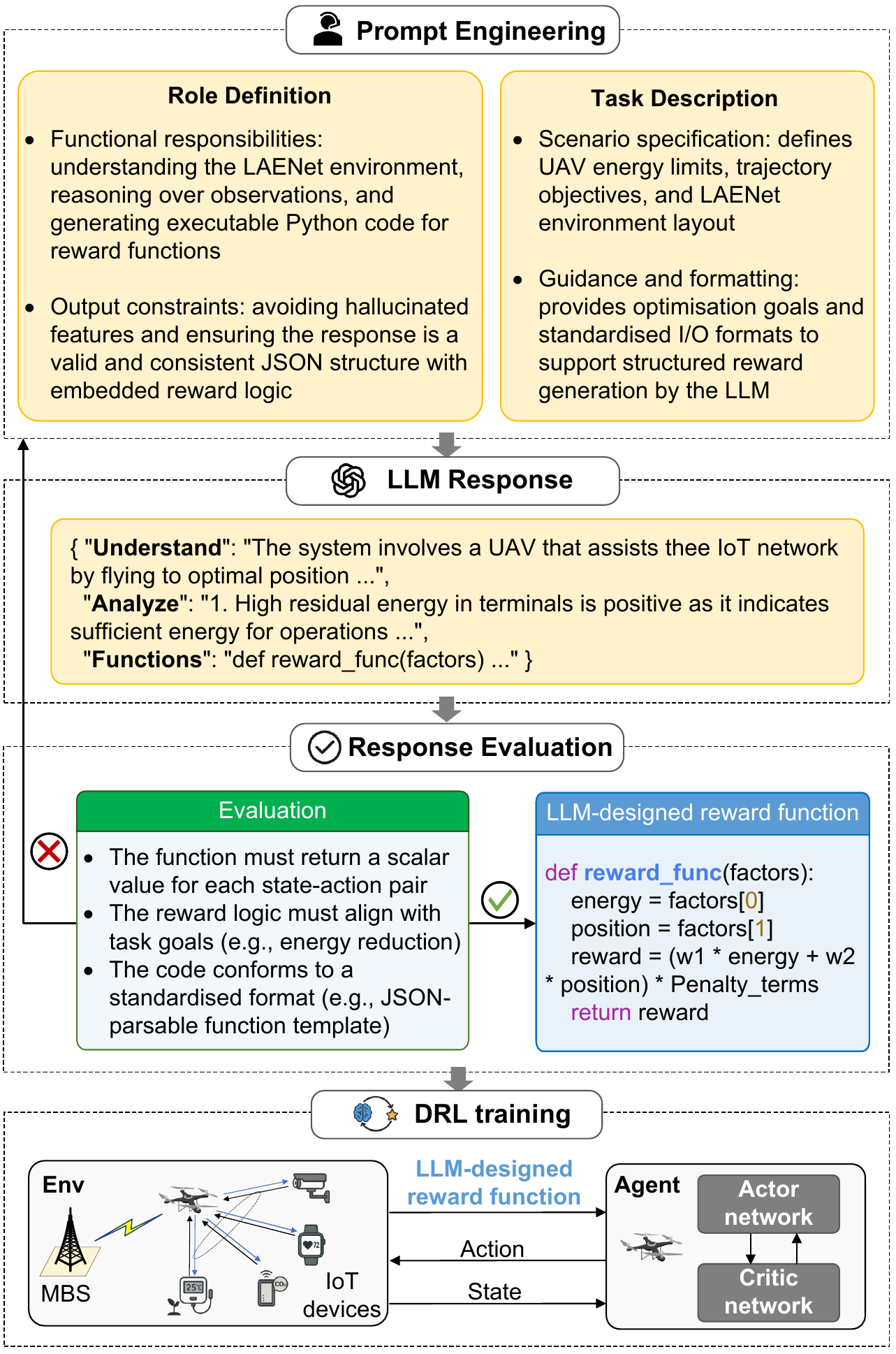}
\caption{LLM-assisted reward design framework for DRL in the LAENet. The framework guides LLMs to generate executable reward functions via prompt engineering. Validated outputs are used to train a DRL agent for UAV energy optimization in low-altitude IoT networks.}
\label{tutorialUAVcasefig}
\end{figure}

\subsection{Case Study: Enhancing DRL for Energy Optimization in LAENet through LLM as reward designer and generator}

\subsubsection{System Description}The case study focuses on the LAENet scenario that consists of multiple IoT devices and a macro base station (MBS) for uplink communication. As illustrated in Fig. \ref{tutorialUAVcasefig}, a UAV serves as a mobile agent tasked with collecting data from distributed Internet of Things (IoT) terminals. These terminals include environmental sensors, health monitors, and surveillance cameras, all deployed across a target area requiring efficient aerial coverage. The UAV’s role is to gather sensory data and relay it back to the MBS, while operating under constraints such as limited onboard energy and time-sensitive service requirements. The DRL agent governs the UAV’s behavior via a policy controller. At each decision time step, the UAV observes the system state (such as terminal locations, data freshness, and energy levels), selects an action accordingly (e.g., trajectory updates and hovering decisions) and receives feedback of reward from the environment.

\subsubsection{Workflow of Using an LLM as a reward designer and generator for DRL} The design of reward functions is a critical factor in determining the convergence speed and performance of DRL agents. To enhance flexibility and contextual alignment, we adopt an LLM-enhanced reward design workflow for energy optimization in LAENet. The workflow comprises several key steps as follows:

{\bf Step 1: Constructing Effective Prompts for the LLM.} To enable the LLM to function as a reliable reward designer, the user needs to formulate a structured prompt, which comprises two essential components:

\begin{itemize}

\item Role Definition: The LLM is explicitly instructed to act as a reward function generator for the RL agent, including understanding the LAENet environment, reasoning based on system observations, and producing executable Python code. Constraints are included to prevent hallucination and to maintain consistency, such as the LLM is instructed to invent features and to output a valid JSON response containing the reward logic.

\item Task Description: The prompt defines the UAV’s operating scenario in LAENet, including energy constraints, trajectory objectives, and environmental layout. The description offers specific guidance on optimization goals (e.g., minimizing propulsion energy while maintaining coverage), allowing the LLM to identify and prioritize meaningful reward factors accordingly. Standardized input and output formats are included to guide structured generation.

\end{itemize}

{\bf Step 2: Generating Candidate Reward Functions via LLM}. Once the prompt is issued, the LLM generates candidate reward functions by combining its logical reasoning abilities with code generation capabilities. Each output includes both a textual explanation and an executable function (e.g., Python code). Due to the stochastic nature of LLM outputs, responses may vary even under similar prompts. Thus, it is necessary to validate the generated reward functions before integrating it into the DRL. Additionally, structured evaluation is essential since LLMs may occasionally produce syntactically correct but semantically invalid outputs.

{\bf Step 3: Evaluating LLM Responses for Validity and Consistency}. Rather than relying on a single output, the proposed scheme prompts the LLM multiple times to generate a diverse set of candidate reward functions. Then, these functions are evaluated against a set of predefined constraints: \begin{itemize}

\item The function must return a scalar value for each state–action pair.

\item The reward logic must align with task goals (e.g., energy reduction).

\item The code conforms to a standardised format (e.g., JSON-parsable function template).

\end{itemize}

Only candidate functions that pass all checks are retained for training the RL agent.

{\bf Step 4: Using the LLM as a Generator to Evaluate Reward Design.} The LLM constructs a semantic proxy of the LAENet environment by generating synthetic samples (i.e., state $s$, action $a$ and reward $R(\cdot)$) that approximate realistic state–action–reward relationships. This process allows \textit{a-priori} evaluation of the reward functions without deploying an actual policy or running RL episodes, effectively reducing the sampling cost and avoiding unnecessary environment interactions. For each candidate reward function $R(\cdot)$, we numerically compute the Lipschitz constant $\mathcal{L}$ over the state space $\mathcal{S}$ as
\begin{equation}
\mathcal{L} = \sup_{s_1,s_2 \in \mathcal{S}, s_1 \neq s_2} 
    \frac{\| R(s_1) - R(s_2) \|_2}{\| s_1 - s_2 \|_2}.
\end{equation}
A smaller Lipschitz constant implies that the reward landscape varies more smoothly with respect to action perturbations \cite{khromov2024some}, leading to better training stability and faster convergence of the DRL agent. Following this principle, the LLM-generated reward functions are ranked according to their estimated Lipschitz constants, and the one with the lowest $\mathcal{L}$ is selected as the final reward design for integration into the RL training process. 

{\bf Step 5: Integrating Expanded Reward Factors for UAV Energy Optimization}. One of the advantages of LLM-designed reward functions is the ability to integrate additional reward components beyond conventional formulations. In the implementation, the manually defined reward is usually designed in the form of $\text{reward} = w \times \text{energy} \times \text{Penalty}$ \cite{9750860}. After using the LLM to design the reward function, a richer reward structure can be inferred and generated by $\text{reward} = \left( w_1 \times \text{energy} + w_2 \times \text{position\_score} \right) \times \text{Penalty}$, where $\text{position\_score}$ reflects how close the UAV is to the centroid of the IoT terminal distribution. This term encourages the UAV to maintain a central position, reducing both travel distance and hovering time. Such enhancements lead to more effective energy use by combining path planning and transmission efficiency via comprehensive and unified reward signals.

\begin{figure}[t]
\centering
\includegraphics[width=1.01\linewidth]{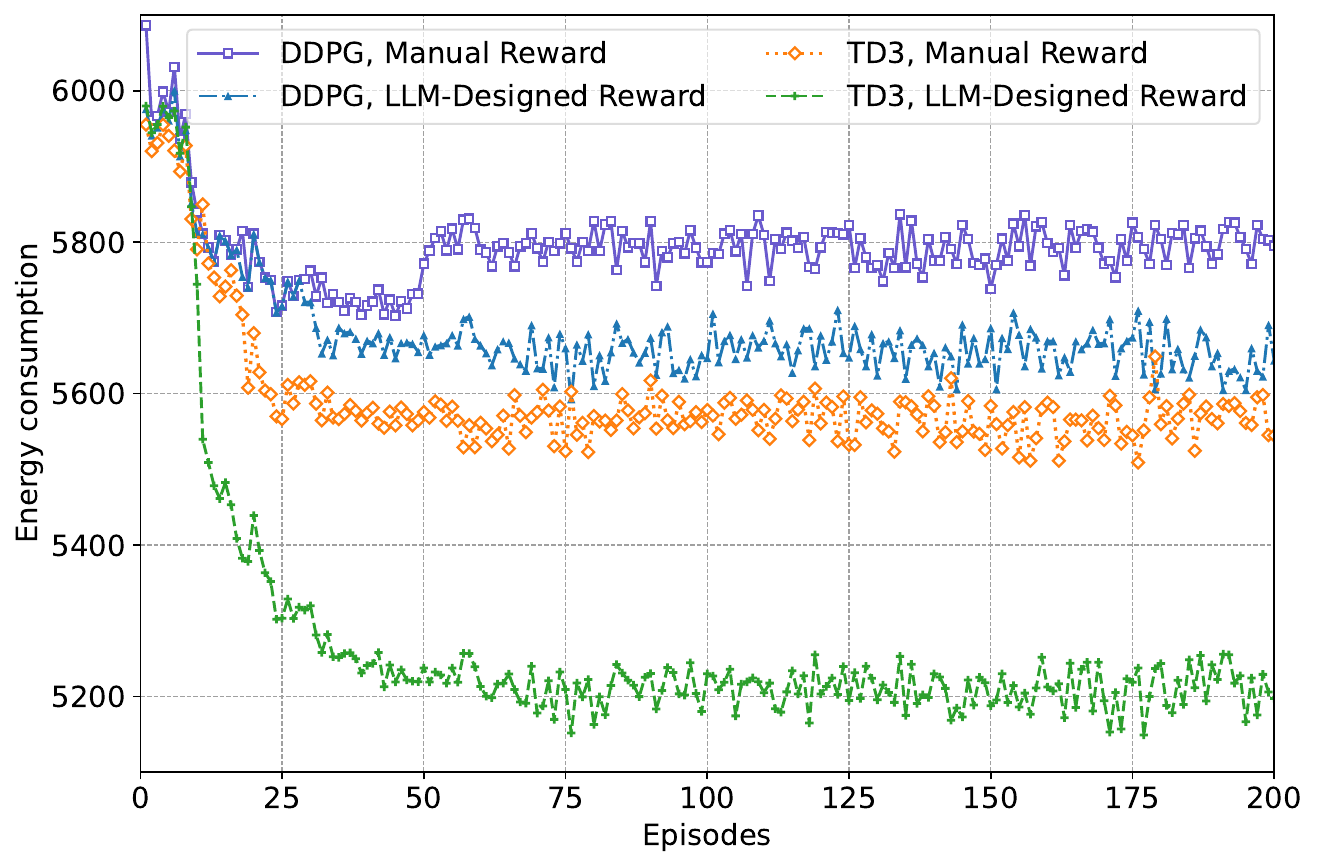}
\caption{Energy consumption across training episodes using manual and LLM-designed reward functions in the LAENet. The results compare DDPG and TD3 algorithms under two reward settings. LLM-designed reward functions lead to lower and more stable energy consumption.}
\label{tutorialUAVcaseexpfig}
\end{figure}

\subsubsection{Numerical Results} Fig. \ref{tutorialUAVcaseexpfig} presents the training performance of two representative DRL algorithms (i.e., DDPG and TD3) under two reward design schemes: conventional manually defined functions and LLM-generated functions. The evaluation metric is the total energy consumption of the UAV across training episodes. From the results, it is evident that incorporating LLM-designed reward functions leads to improvements in energy efficiency across two algorithms. The TD3 agent equipped with the LLM-derived reward achieves the lowest final energy consumption, showing a reduction of up to 6.8\% compared to its manually designed counterpart. The performance gain stems from the LLM’s ability to generate reward structures that better reflect the optimization objectives of LAENet. By reasoning over UAV dynamics, positional relevance, and energy constraints, the LLM formulates reward signals that implicitly guide the agent toward shorter flight paths, reduced hovering time, and more efficient communication scheduling. Thus, the simulation results demonstrate the effectiveness of LLM-assisted reward design in enhancing the decision-making capabilities of DRL agents for the LAENet.

{\bf Lesson Learned.} The case study demonstrates that serving LLMs as the reward designer and generator can improve the efficiency and stability of UAV energy optimization. The effectiveness of this approach depends on grounding LLM reasoning in domain knowledge of low-altitude communication and mobility, ensuring that the generated reward function aligns learning objectives with mission-level constraints. Meanwhile, the LLM synthesizes state–reward samples to further accelerate and ensure the stability and feasibility of reward design. Future research can leverage LLMs’ abilities to interpret heterogeneous data streams and reason over sequential context, allowing RL agents to translate multimodal data (such as raw imagery, sensory data, and mission updates) into comprehensive state representations for RL. Moreover, step-by-step reasoning mechanisms such as CoT prompting can be further explored to infer mission intent, operational constraints, and decision making for optimization in the LAENet.

\section{LLM-enhanced RL for Vehicular Networks}

\subsection{Background and Motivation}

\begin{figure*}[!t]
\centering
\includegraphics[width=6.8in]{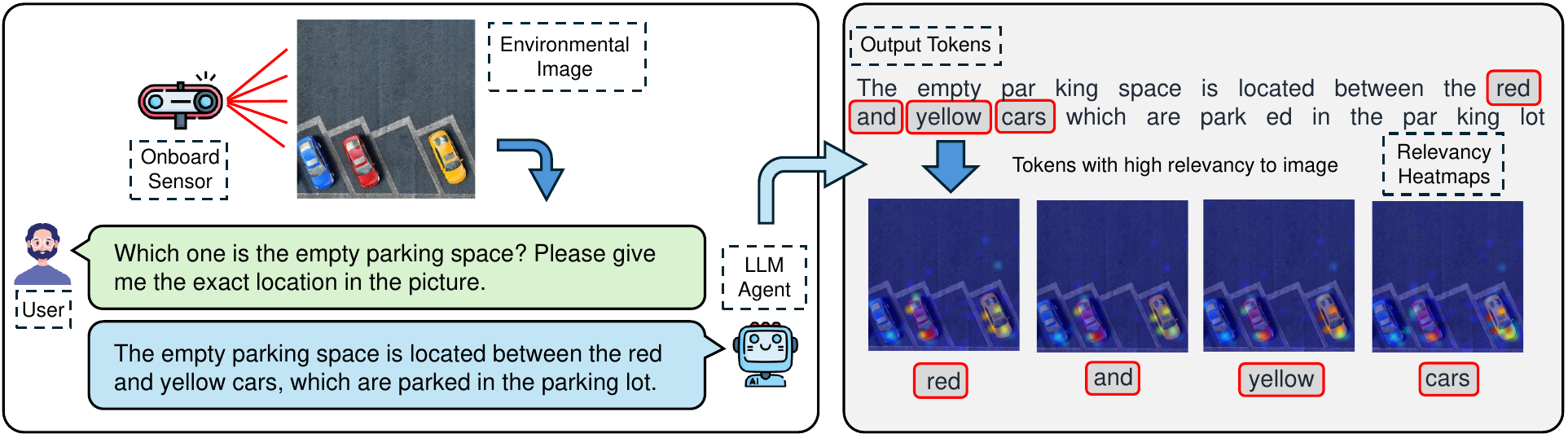}
\caption{Illustration of the LLM as state perceiver in an embodied-AI vehicle. LLAVA processes onboard camera images to identify semantic details, such as the empty parking space between the red and yellow cars. The relevancy heatmaps demonstrate accurate cross-modal grounding between visual regions and textual tokens, confirming the model’s ability to extract and interpret environment-aware semantic information.}
\label{vehicularcase}
\end{figure*}

With the advent of the 6G era, the Internet of Vehicles (IoV) is expected to deliver orders-of-magnitude increases in traffic and connectivity, enabling real-time navigation, perception, and autonomous decision-making at scale \cite{9808399}. Vehicular networks sit at the core of this vision by linking vehicles and roadside infrastructure, where the proliferation of onboard and roadside sensors creates a continuous stream of multimodal data that must be processed and exchanged under tight latency and bandwidth budgets \cite{jiang2018low}. DRL can serve the latter by learning policies that continuously reconfigure channel usage, power, and scheduling from feedback \cite{8714026}. Although traditional DRL is effective for object detection and path planning in static settings, it struggles to sustain accuracy and responsiveness when exposed to fast topology changes, heterogeneous scenes, and concurrent tasks \cite{10660494}. Thus, it is crucial to develop a scheme that can turn rich sensor feeds into compact and decision-useful states while adapting online to non-stationary wireless conditions.

LLMs can serve the former by extracting salient semantics from images and videos and converting them into structured, text-level representations that are cheaper to transmit and easier to fuse \cite{Wang2024,wu2024next}. In this case, LLMs can serve as state perceivers for vehicular networks. Specifically, the LLM operates as a multimodal front end that distills raw sensor streams into compact semantic descriptors for RL state representation. Coupled with DRL, which optimizes actions from these enriched states, this integration links high-level semantic understanding with low-level control, improving sample efficiency, convergence behavior, and end-to-end QoE in dynamic vehicular environments.

\subsection{Case Study: LLM as a State Perceiver for DRL in quality of experience (QoE) Maximization }

\subsubsection{System Description} We consider a cellular-based vehicular network deployed in an urban area, where a base station (BS) serves a fleet of embodied-AI vehicles. Association follows the strongest RSSI at the BS \cite{10041763}. Communication occurs over two link types: V2I links carry high-priority or aggregated reports to the BS, while V2V links enable nearby vehicles to exchange information locally, relieving backhaul load and latency \cite{9808399}. Each vehicle runs an onboard multimodal front end powered by an LLM to turn camera feeds into structured semantic messages (e.g., objects, counts, relations, and traffic cues). These messages are compacted by a semantic encoder and then prepared by a channel encoder for over-the-air delivery. Because V2V links may reuse V2I subbands, cochannel interference arises and the effective SINR on each link becomes time-varying. At the receiver, channel and semantic decoders reconstruct the message and its meaning. Reconstruction quality is tracked using cosine similarity between BERT embeddings, complemented by mutual information to gauge how well semantics survive the wireless hop \cite{10746594}.

The system objective is to maximize a QoE metric inspired by the Weber–Fechner law, jointly reflecting semantic fidelity and resource usage \cite{10654734}. Operational constraints include binary V2V–V2I sharing choices, one-to-one subband assignments, admissible symbol lengths, SINR thresholds for reliable reception, and power bounds for each transmitter. Vehicles act as embodied AI agents that tune communication decisions online. A DRL controller adapts channel reuse, transmit power, and the number of semantic symbols carried per report. The result is a closed-loop stack where LLM-based perception supplies compact, task-aware state to the RL controller, which in turn steers spectrum and payload to sustain high QoE under mobility and interference.

\subsubsection{Workflow of LLM as State Perceiver for DRL}
To tighten the perception–decision loop in vehicular networks, we adopt a vision–language model (LLAVA) \cite{Xie_2023_CVPR} distills raw images into compact, task-aware semantics that feed the DRL controller. The workflow proceeds as follows.

\textbf{Step 1: Capture and pre-process multimodal observations.}
Each embodied-AI vehicle collects camera frames $\mathbf{I}_i$ together with basic link/context metadata. Images are normalized and fed to the vision front end so that only the information useful for downstream communication and driving decisions is retained.

\textbf{Step 2: Visual feature extraction with CLIP-style encoder.}
LLAVA employs a contrastive language–image pre-training visual encoder $g(\cdot)$ to obtain compact features
$\mathbf{Z}_i = g(\mathbf{I}_i)$,
which summarize salient scene entities (i.e., vehicles, pedestrians, lanes, and signals) and their coarse relations.

\textbf{Step 3: Visual–language alignment for semantic generation.}
The visual features extracted by the encoder are subsequently aligned with the linguistic embedding space through a trainable linear projection, expressed as $\mathbf{E}_i = \mathbf{W}\mathbf{Z}_i$. This mapping establishes a shared representation domain that bridges visual and textual modalities. Building on this alignment, the LLAVA model generates the corresponding semantic representation $\mathbf{M}_i = \text{LLAVA}(\mathbf{I}_i;\theta_i)$, where $\theta_i$ denotes the model parameters. The resulting $\mathbf{M}_i$ encapsulates high-level contextual descriptions of the observed scene (such as object identities, spatial relations, and environmental attributes) that serve as compact, information-rich inputs for downstream transmission and decision-making processes.

\textbf{Step 4: Scenario adaptation via lightweight fine-tuning.}
To adapt the pretrained LLAVA model to the specific characteristics of real-world driving environments, the model is fine-tuned on vehicular data to minimize semantic mismatch between ground-truth semantics and predictions. This step improves robustness to urban lighting, occlusions, and viewpoint changes while keeping the front end computationally tractable on vehicle/edge platforms.

\textbf{Step 5: Sentence-level encoding and quality tracking.}
For stable ingestion by the RL agent, $\mathbf{M}_i$ is encoded by a BERT model $B(\cdot)$ into a sentence-level vector. Reconstruction fidelity along the semantic communication pipeline is monitored by cosine similarity, which later participates in QoE calculation and thus links perception quality to control objectives.

\textbf{Step 6: State assembly for the DRL backbone.}
The RL state aggregates (i) semantic descriptors from $B(\mathbf{M}_i)$ (e.g., counts, occupancy relations, and traffic cues) and (ii) communication-side observables, such as subband reuse indicators and interference/SINR statistics. As a result, a compact and task-relevant representation is produced that replaces pixel-heavy inputs and exposes high-level structure to the agent.

\textbf{Step 7: DRL-driven communication control.}
Given the LLM-enhanced state, a PPO agent with generalized advantage estimation selects actions over channel reuse, transmit power, and the number of semantic symbols per report. Decisions are evaluated by a QoE objective inspired by the Weber–Fechner law, balancing semantic fidelity and resource usage under mobility and cochannel interference.

\textbf{Step 8: Closed-loop update and continual refinement.}
Rewards derived from QoE and constraint satisfaction (e.g., SINR and similarity thresholds) drive policy updates. In parallel, monitoring signals (e.g., drops in $\xi$ under specific scenes) can trigger periodic fine-tuning of LLAVA on recent samples, keeping the perceiver aligned with operation conditions and sustaining end-to-end convergence and stability.

\subsubsection{Numerical Results}

\begin{figure}[t]
\centering
\includegraphics[width=1.0\linewidth]{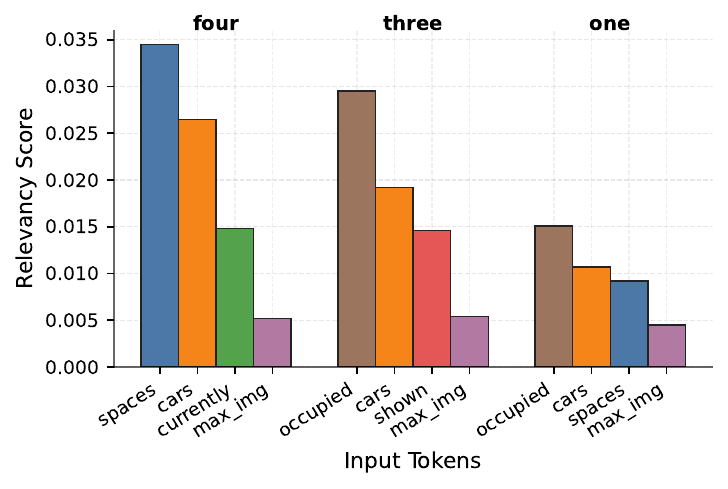}
\caption{Input–output text relevancy analysis of LLAVA in semantic perception. The bar charts illustrate the relevancy scores between input tokens and output tokens when LLAVA describes the parking scenario. Tokens such as spaces, occupied, and cars exhibit the highest contribution to generating the output tokens ``four", ``three", and ``one", respectively.}
\label{vehicularcaseexpfig1}
\end{figure}

\begin{figure}[t]
\centering
\includegraphics[width=1.0\linewidth]{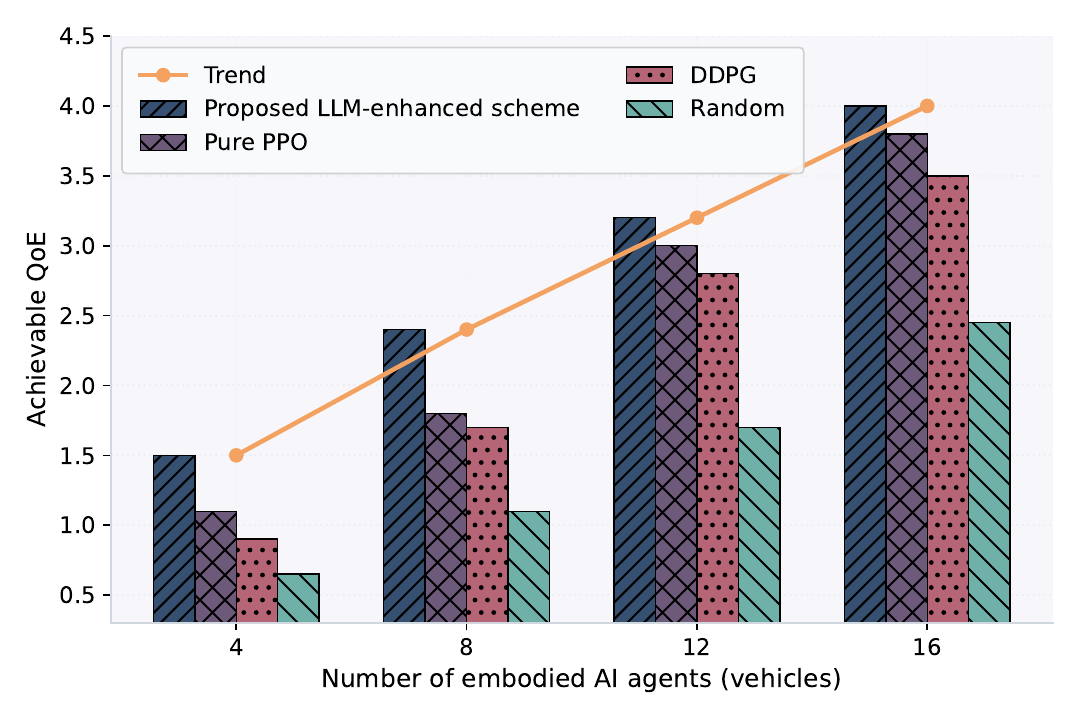}
\caption{Achievable QoE versus the number of LLM agents.}
\label{qoefigvcase}
\end{figure}

As shown in Fig. \ref{vehicularcaseexpfig1}, LLAVA effectively identifies critical contextual cues such as the number of available parking spaces, their occupancy status, and the spatial relationship between vehicles. In this scenario, the system correctly infers that four parking spaces are present, three are occupied, and one remains vacant. Compared to transmitting the original image (approximately 614 KB), conveying the extracted textual description (around 12.1 KB) reduces the transmission bandwidth by nearly 98\%, demonstrating the advantage of semantic-level communication in bandwidth-constrained vehicular networks.

As shown in Fig. \ref{qoefigvcase}, the proposed LLM-enhanced scheme consistently achieves the highest QoE across all network scales. The system exhibits a 36\% QoE improvement over DDPG at eight vehicles and maintains steady scalability with 61.4\%, 31.9\%, and 25.2\% incremental gains when the number of vehicles increases from 4 to 8, 8 to 12, and 12 to 16, respectively. These results highlight that the LLM-enhanced state representation contributes directly to the observed performance improvement, supporting faster learning and higher-quality decision making in large-scale vehicular networks.

{\bf Lesson Learned.} Integrating LLM-enhanced RL into vehicular networks demonstrates that semantic understanding can fundamentally reshape communication and control in bandwidth-limited environments. By extracting and transmitting high-level scene semantics rather than raw sensory data, LLMs reduce communication overhead while preserving decision-critical information. This integration also illustrates the promise of hierarchical cooperation in LLM-enhanced RL for vehicular networks: LLMs provide semantic abstraction and intent reasoning, while RL agents formulate fine-grained decision optimization. Future work should prioritize lightweight multimodal compression, semantic calibration, and reliability assessment to ensure that LLM-enhanced RL can achieve scalable, robust, and trustworthy vehicular intelligence.

\section{LLM-enhanced RL for Space-air-ground Integrated Network}

\subsection{Background and Motivation}

Space–air–ground integrated network (SAGIN) is a next-generation architecture that couples spaceborne, airborne, and terrestrial segments into a unified system for wide-area connectivity \cite{10670196}. In this paradigm, low Earth orbit (LEO) constellations furnish global coverage and low-latency backhaul, high-altitude platforms (HAPs) at stratospheric altitudes operate as quasi-stationary relays and traffic aggregators, while ground infrastructures provide access and edge computation \cite{widiawan2007high}. A hybrid free-space optical (FSO)/radio frequency (RF) stack naturally aligns with this multi-tier design, where FSO links are suitable for satellite-to-HAP segments due to high bandwidth and license-free spectrum at high altitudes, whereas RF links offer robust HAP-to-ground access under cloud, fog, and precipitation \cite{10670196}. Despite these advantages, SAGIN presents tightly coupled design challenges. First, the fast orbital motion of LEO satellites continuously reshapes the topology, making handover planning and inter-segment continuity a critical issue. Second, heterogeneous FSO and RF links introduce cross-tier dependencies in resource allocation. Decisions on the satellite–HAP segment affect the feasible operating region of the HAP–ground segment. Third, although HAPs are quasi-stationary, slow drifts caused by changing environmental conditions perturb link geometry and channel statistics. These effects lead to non-convex objectives, time-coupled constraints, and a large combinatorial action space, where handover, user selection, and subcarrier/power scheduling must be coordinated across tiers \cite{9738819}.


To cope with these dynamics, many studies leverage RL as a model-free controller that adapts online to mobility and channel fluctuations \cite{10299604,9329087,9372298}. Nevertheless, classical RL still faces practical limitations in SAGIN such as limited generalization to unseen traffic/topology regimes and brittle performance under non-stationary channels when training with fixed hyperparameters \cite{NEURIPS2020_5a751d6a,moerland2023model}. These issues are amplified by the hybrid FSO/RF coupling and the visibility-constrained satellite selection space. Recent advances in LLMs provide a transformative opportunity to enhance DRL in SAGIN. An LLM can function as a decision guider that oversees and guides the learning dynamics of the DRL agent \cite{10638533,10648594}. Leveraging reasoning and contextual understanding capabilities, the LLM can interpret training feedback such as reward trajectories, convergence patterns, and exploration progress to infer how the learning process should adapt over time. This integration paradigm enables DRL to combine low-level environmental interaction with high-level adaptive oversight. Building on this perspective, the following subsection presents representative case studies of LLM-enhanced RL for SAGIN.

\subsection{Case Study: LLM as a decision guider for DRL in downlink throughput maximization}

\begin{figure}[t]
\centering
\includegraphics[width=1.0\linewidth]{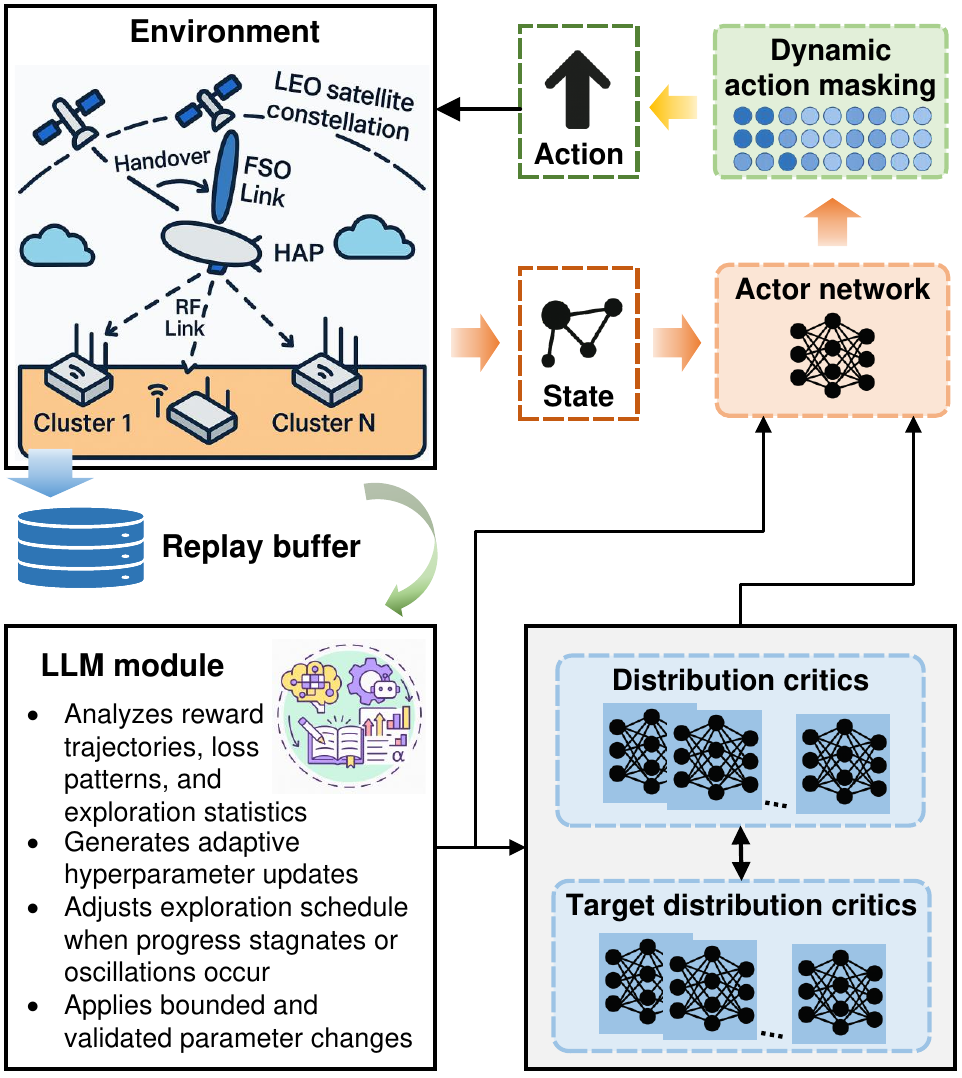}
\caption{Framework of the proposed LLM-guided TQC for hybrid satellite downlink optimization. The LLM module serves as a meta-controller that analyzes learning behavior and adaptively tunes hyperparameters to stabilize and accelerate training.}
\label{sagincasefig}
\end{figure}

\subsubsection{System Description}As shown in Fig. \ref{sagincasefig}, the considered system consists of a three-tier space–air–ground downlink architecture, where a constellation of LEO satellites provides high-capacity backhaul to a HAP hovering in the stratosphere, which in turn distributes data to multiple ground user clusters. The system adopts a hybrid FSO and RF communication framework: the satellite–to–HAP link uses FSO for its large bandwidth and interference-free propagation above the clouds, while the HAP–to–ground segment employs RF transmission to maintain reliable connections under variable weather conditions. Within each time slot, the network dynamically selects one visible LEO satellite to establish the FSO backhaul and simultaneously allocates limited RF subcarriers among ground clusters according to real-time channel quality. The optimization goal is to achieve joint coordination between satellite handover and network resource allocation, i.e., to maximize the overall downlink throughput from space to ground while minimizing frequent satellite handovers that can cause service interruptions and control overhead.

\subsubsection{Workflow of Using LLM as a Decision Guider for DRL}

To make RL robust under the non-stationary and hybrid FSO/RF conditions of a space–air–ground system, we function the LLM \emph{above} as the guider for the RL agent in policy learning. It periodically inspects training information and decides \emph{how the guider should learn next}. The workflow is organized into the following steps.

\textbf{Step 1: Define the LLM’s role, guardrails and training information.}
Configure the LLM to act as a decision guider over the DRL backbone (TQC)\footnote{TQC \cite{pmlr-v119-kuznetsov20a} is an advanced variant of SAC based on distributional RL. Instead of estimating the expected Q-value, TQC models the full return distribution using multiple quantile critics and mitigates overestimation bias by truncating the highest quantiles during value updates, thereby achieving more stable and reliable policy learning in high-variance environments.}. Its scope is restricted to learning dynamics (such as stability, progress, exploration), not task actions. We whitelist adjustable hyperparameters (e.g., learning rate, entropy/temperature $\alpha$, truncation level $k$ in TQC, batch size, target-network update rate, exploration decay) and set intervention cadence and safety constraints (i.e., min/max ranges, maximum step sizes). On a fixed schedule, the training loop assembles a structured information package including recent reward trajectory, actor/critic loss traces, entropy/exploration statistics, and a normalized training-progress indicator. 

\textbf{Step 2: Prune the action space with dynamic action masking.}
A deterministic feasibility layer masks out invalid actions (e.g., satellites that are not currently visible), which ensures compute is focused on meaningful choices and that any exploration encouraged by the LLM happens within the feasible set.

\textbf{Step 3: Calibrate exploration with an interpretable schedule.}
To avoid both aimless exploration and premature exploitation, the agent follows an episode-dependent exploration schedule:
\begin{equation}
\epsilon(e)=\max\!\left(\epsilon_0\!\left(1-\frac{e}{e_{\text{decay}}\cdot E}\right),\,0\right),
\end{equation}
where $\epsilon_0$ is the initial noise level, $e$ is the current episode, $E$ is the total episode budget, and $e_{\text{decay}}$ sets how quickly exploration tapers off. The LLM may later suggest gentle, bounded adjustments to this schedule (e.g., slowing the decay when progress plateaus).

\textbf{Step 4: Prompt the LLM for adjustments.}
At each intervention, the system presents a structured summary including current hyperparameters, recent rewards, and progress marker. The LLM outputs a set of small and coordinated hyperparameter adjustments that are consistent with the diagnosed learning phase: \begin{equation}
\Theta_{e+\Delta e}=\mathcal{F}_{\text{LLM}}\!\left(\Theta_e,\ \mathcal{H}_e,\ \mathcal{P}_e\right),
\end{equation}
where $\Theta_e$ is the current hyperparameter set enhanced bu an LLM $\mathcal{F}_{\text{LLM}}$, $\mathcal{H}_e$ summarizes recent rewards, and $\mathcal{P}_e$ encodes training progress. For instance, when oscillations are detected in the reward curve, the LLM suggests reducing the learning rate to stabilize convergence; conversely, when training enters a plateau phase, it recommends increasing the entropy coefficient to enhance exploration. 

\textbf{Step 5: Maintain Effective Adaptation and Continuous Feedback.}
Before deployment, a local validator clamps every suggested change into certified ranges and rejects out-of-scope edits. Rate-limit rules cap per-interval movement to avoid instability (e.g., no large jumps in learning rate or $\alpha$). Validated updates are applied for the next window. Fresh training information reflects the consequence of prior advice and becomes the next input to the LLM. This closes a feedback loop where the LLM continually refines the learning process in response to the non-stationary environment characterized by time-varying channel conditions, satellite mobility, and HAP dynamics.

\textbf{Step 6: Log rationales and handle failure modes.}
Each LLM recommendation is stored with a short rationale (e.g., oscillatory rewards mean the lower step size, slightly increase entropy). If performance degrades beyond a tolerance, an automatic rollback restores the last stable configuration; the next prompt asks the LLM for a corrective countermeasure.

\subsubsection{Numerical Results} Numerical results show that compared to the classical TQC, the proposed LLM-guided TQC improves approximately 1.5\% to 2\% higher steady-state rewards, demonstrating the benefit of LLM-driven hyperparameter adaptation and invalid-action elimination. Against the baselines SAC, PPO, TD3, and DQN, the proposed method achieves around 3\% to 12\% reward improvement and converges within fewer than 400 episodes, while other algorithms exhibit slower stabilization and larger reward fluctuations. These results verify that incorporating LLM for decision guiding effectively enhances learning stability and adaptability in dynamic satellite communication environments.

{\bf Lesson Learned.} The application of LLM-enhanced RL in SAGIN achieves improved coordination across heterogeneous links by utilizing the LLM as a decision guider to adapt hyperparameters and exploration schedules of RL. Such oversight is achieved through the LLM’s capability for reasoning over sequential patterns and interpreting training feedback, enabling faster convergence for the RL agent in policy learning. However, it is important to note that when LLMs are used to enhance RL through parameter tuning, the embedding of expert-level domain prior knowledge is essential to ensure that the tuning decisions are meaningful and robust. Therefore, developing domain-specific LLMs trained on communication and network optimization corpora (beyond SAGIN) represents a promising research direction.






\section{Future Research Directions}

Despite the promising progress of integrating LLMs into RL for wireless networks, this emerging paradigm remains in its infancy. The fusion of linguistic cognition and autonomous decision-making opens a wide spectrum of opportunities that demand deeper theoretical insights, efficient system design, and reliable deployment practices. In this section, we outline several promising future research directions on the LLM-enhanced RL.

\subsection{Theoretical Foundations and Performance Guarantees}

While LLM-enhanced RL has demonstrated remarkable adaptability and generalization, its theoretical underpinnings remain largely unexplored. This is primarily because the LLM is often treated as a closed-box component \cite{Wang2024}, making it unclear how LLMs theoretically influence the reinforcement learning process. Future research needs to bridge this gap by developing rigorous mathematical tools to analyze how semantic reasoning introduced by LLMs affects the convergence, stability, and optimality of RL \cite{10.1145/3626772.3657767}. Promising directions include (i) quantifying the sensitivity of LLM-generated states and rewards, and (ii) characterizing the uncertainty arising from LLM reasoning or hallucinations to ensure stable policy updates. Such theoretical progress is essential for transforming LLM-enhanced RL from empirical closed-box integration into a principled and verifiable learning paradigm \cite{NEURIPS2022_266c0f19}.

\subsection{Lightweight and Edge-Deployable LLM-Enhanced RL Architectures}
The computation-intensive nature of LLMs poses challenges for their real-time integration with RL in wireless systems \cite{10.5555/3600270.3602446}. Future research should pursue lightweight, modular, and domain-adaptive architectures that can retain the semantic reasoning capabilities of LLMs while satisfying strict energy, latency, and storage constraints \cite{10648594}. Feasible research directions include knowledge distillation \cite{gou2021knowledge} and parameter-efficient fine-tuning \cite{hu2022lora} to compress foundation models for efficient inference. In addition, edge–cloud collaborative inference frameworks can be developed to offload heavy semantic reasoning to nearby edge servers while retaining fast decision loops locally \cite{10759588}. Such lightweight deployment is crucial for enabling efficient LLM-enhanced RL in practical applications of next-generation wireless networks.

\subsection{Security and Trustworthy LLM-Enhanced RL Systems}

Security assurance becomes paramount as LLMs participate in network control and resource allocation within wireless systems \cite{10614634}. Potential risks such as prompt injection, model hallucination, and adversarial manipulation may lead to catastrophic misallocations or service disruptions \cite{liu2024exploring,10648594,ali2023leveraging}. Future work should focus on establishing trustworthy LLM-enhanced RL pipelines. For example, designing robust prompt architectures and adversarial defense mechanisms can help resist malicious or misleading inputs. Abnormal responses caused by hallucinations during the RL agent’s policy execution can be avoided by incorporating consistency verification mechanisms. These efforts will be essential for ensuring regulatory compliance and building trust in high-stakes wireless infrastructures such as SAGIN and LAENet \cite{10299604,cai2025large}.

\subsection{Multi-Agent and Collaborative LLM-RL Coordination}

Next-generation wireless systems involve multiple agents (such as UAVs, vehicles, base stations, and edge devices) to achieve collaborative intelligence \cite{9372298}. Extending LLM-enhanced RL to MARL or collaborative learning scenarios introduces new research challenges \cite{10638533}. For example, using a single type of LLM to interpret, negotiate, and summarize the intents of distributed agents may face cognitive limitations \cite{kannan2024smart}. How can communication efficiency and policy consistency be ensured when multiple LLMs operate concurrently under bandwidth constraints \cite{10.1145/3712678.3721880}? Therefore, integrating multiple types of LLMs into RL agents responsible for different tasks to construct a collaborative decision-making framework represents a novel research direction.

\subsection{Wireless Network-Specific LLM Pretraining for RL}

Generic foundation models trained on web-scale corpora may lack awareness of the physical, statistical, and semantic characteristics of wireless environments \cite{10614634}. Hence, a key frontier lies in developing domain-specialized LLMs for wireless networks \cite{gajjar2025oransight}. These LLMs can be pretrained on multimodal datasets including communication logs, signal features, network management commands, and UAV control scripts \cite{lotfi2025oran,yang2025llmkey,xu2025scalable}. Such specialization will endow LLMs with deeper domain reasoning and understanding, thereby improving interpretability and decision accuracy when integrated with RL agents.

\section{Conclusion}

This tutorial has provided a comprehensive overview of the emerging paradigm of LLM-enhanced RL for wireless networks. By exploring how LLMs can be incorporated into the RL pipeline, we have shown LLM's superior abilities to overcome challenges of the classical RL. We propose the taxonomy, which covers the roles of LLMs as state perceiver, reward designer, decision-maker, and generator for RL agent, and offers the latest overview and unique insights into LLM-enhanced RL in wireless systems. Through multiple case studies in the LAENet, vehicular networks, and SAGIN, we demonstrated that LLM-enhanced RL can significantly improve adaptability, stability, and efficiency in dynamic wireless environments. Looking forward, several open challenges remain to be addressed, which will be key issues to realize the full potential of LLM-enhanced RL in wireless networks.


\bibliographystyle{IEEEtranN}
\bibliography{cly, hyx, content/reference}

\vfill

\end{document}